\documentclass[reprint,NumberedRefs]{JASA}
\usepackage{stackrel}
\usepackage{comment}
\usepackage{tabularx}
\usepackage{siunitx} % for nicely formatted numbers and units

\usepackage{placeins}

\begin{document}
%% the square bracket argument will send term to running head in
%% preprint, or running foot in reprint style.

\title[]{Enhancing the analysis of murine neonatal ultrasonic vocalizations: Development, evaluation, and application of different mathematical models}
%The title must be in sentence case (i.e., lower case with only the first word and proper names capitalized)
% ie
%\title[JASA/Sample JASA Article]{Article title should be less than 17 words, no acronyms}

\author{Rudolf Herdt}
\affiliation{Center for Industrial Mathematics,  University of Bremen, Bremen 28334, Germany}
\author{Louisa Kinzel}
\affiliation{Center for Industrial Mathematics,  University of Bremen, Bremen 28334, Germany}
\author{Johann Georg Maaß}
\email{johann.maass@med.uni-heidelberg.de}
\affiliation{Institute of Human Genetics,  University of Heidelberg, Heidelberg 69120, Germany}
\affiliation{Also at: Interdisciplinary Neurobehavioral Core, University of Heidelberg, Heidelberg, 69120,
Germany}
\author{Marvin Walther}
\affiliation{Institute of Electrodynamics and Microelectronics,  University of Bremen, Bremen 28334, Germany}
\author{Henning Fröhlich}
\affiliation{Institute of Human Genetics,  University of Heidelberg, Heidelberg 69120, Germany}
\author{Tim Schubert}
\affiliation{Institute of Human Genetics,  University of Heidelberg, Heidelberg 69120, Germany}
\author{Peter Maass}
\affiliation{Center for Industrial Mathematics,  University of Bremen, Bremen 28334, Germany}
\author{Christian Patrick Schaaf}
\affiliation{Institute of Human Genetics,  University of Heidelberg, Heidelberg 69120, Germany}

%ORCID
% PM https://orcid.org/0000-0003-1448-8345
% CPS https://orcid.org/0000-0002-2148-7490
% HF
% JM https://orcid.org/0009-0006-2069-5470
% MW https://orcid.org/0009-0002-8047-219X
% LK https://orcid.org/0000-0002-0271-175X
% RH 0009-0003-7537-3645
% TS 0009-0003-1696-4402

% ie
%\author{Author One}
%\author{Author Two}
%\author{Author Three}

% ie
%\affiliation{Humangenetics,  University Heidelberg, Heidelberg, State ZipCode, Germany}

%\altaffiliation{}

% may be added after \author{}, ie
% \altaffiliation{Also at: Department1,  University1, City, State ZipCode, Country.}

%% for corresponding author

%% For preprint only,
%  optional, if you want want this message to appear in upper right corner of title page
% \preprint{}

%ie
%\preprint{Author, JASA}

% optional, if desired:
%\date{\today}

\begin{abstract}

Rodents employ a broad spectrum of ultrasonic vocalizations (USVs) for social communication. As these vocalizations offer valuable insights into affective states, social interactions, and developmental stages of animals, various deep learning approaches have aimed to automate both the quantitative (detection) and qualitative (classification) analysis of USVs. Here, we present the first systematic evaluation of different types of neural networks for USV classification. We assessed various feedforward networks, including a custom-built, fully-connected network and convolutional neural network, different residual neural networks (ResNets), an EfficientNet, and a Vision Transformer (ViT). Paired with a refined, entropy-based detection algorithm (achieving recall of \(94.9\,\mathrm{\%} \) and precision of \(99.3\,\mathrm{\%} \)), the best architecture (achieving \(86.79\,\mathrm{\%} \) accuracy) was integrated into a fully automated pipeline capable of analyzing extensive USV datasets with high reliability. Additionally, users can specify an individual minimum accuracy threshold based on their research needs. In this semi-automated setup, the pipeline selectively classifies calls with high pseudo-probability, leaving the rest for manual inspection. Our study focuses exclusively on neonatal USVs. As part of an ongoing phenotyping study, our pipeline has proven to be a valuable tool for identifying key differences in USVs produced by mice with autism-like behaviors.

%193words
%200 words max
%no citations

%Category for JASA: SIGNAL PROCESSING IN ACOUSTICS

%Keywords for publication
% Ultrasonic vocalization
% Neural networks
% Explainable artificial intelligence
% Binary Classification
% Multiclass Classification
% Spectral Entropy
% Deep Learning
% Audio classification
% \\

\end{abstract}

%% pacs numbers not used

\maketitle

\section{Introduction}
\label{sec:Introduction}

Ultrasonic vocalizations (USVs) in rodents are calls above our hearing range (\textgreater \(20\,\mathrm{kHz}\)) which are produced by rodents and play an important role in the social interaction of the animals. Mice, in particular, emit a rich variety of USVs that differ in length, frequency, and modulation. Both the quantity and quality of these calls can provide insights into well-being, development, and sociability.\cite{Portfors2007-kn,Peleh2019-ic,Scattoni2008-vp} The vocalizations of mice evolve throughout development, following an ontogenetic profile.\cite{Elwood1982-vo} Anomalies in USVs have been described in genetic mouse models for autism, schizophrenia, and Down syndrome.\cite{Ey2013-sm, Scattoni2009-zx, Holtzman1996-vf} Due to its high informative value, investigation of call behavior has sparked research interest and has been established as a reliable tool in behavioral neuroscience and pharmacology.\cite{Dirks2002-hy}

While the recording of USVs is a simple and fast procedure, until now, analyzing the recordings has been very labor-intensive, thereby restricting the scientific potential of the experiment. Conventional manual quantification (call detection) and qualitative analysis (call classification) can take up to 2 hours for a 5-minute-long recording. In addition, manually analyzed datasets show low interrater reliability.

New mathematical deep learning models now make it possible to both detect and classify calls, hence speeding up the process and improving the reliability significantly. So far, most efforts have focused on automating the analysis of USVs emitted by rats or adult mice in the context of both affective and aversive states.\cite{Coffey2019-ve,Pessoa2022-sy} In this paper, we focus specifically on neonatal USVs produced upon separation from the litter. 

Studying neonatal USVs in mice offers a unique opportunity to investigate the early developmental aspects of acoustic communication deficits and provides valuable insights into the underlying mechanisms of autism-like behavior.\cite{Takumi2020-nc}
Pups separated from the mother will produce a large quantity and variety of USVs in order to elicit retrieval by the mother.\cite{Ehret2005-ns} Quantifying and classifying these calls are the subject of many neurobehavioral studies. The diverse nature of the calls combined with inevitable background noise complicates the analysis. Even previous advanced models, as reviewed by Pessoa et al. in 2022, face challenges in achieving high values for both recall and precision.\cite{Pessoa2022-sy} The subsequent classification of neonatal USVs is especially difficult as the distribution of calls changes over time.\cite{Peleh2019-ic} Different groups have tried to use a variety of mathematical models and forms of neural networks (threshold-based, fully connected, CNN, recurrent, etc.) to detect and classify calls.\cite{Pessoa2022-sy,Goussha2021-wm,Coffey2019-ve, De_Chaumont2021-ry, Fonseca2021-qr} But different types of data require different types of algorithms. And the question of which network type is best used for the classification of calls is yet to be answered.

It was our goal to develop a fully automated pipeline for the analysis of neonatal USVs. In the first step, we developed an algorithm that can reliably detect USVs in recordings. For the subsequent classification, we built and tested different types of neural networks to determine which network is best suited for the task. In the end, the algorithms for the detection and the classification were concatenated to a comprehensive pipeline that can automatically quantify and classify calls in recordings of pups, using advanced mathematical models.

In this paper, we present the first systematic evaluation of different neural networks for USV classification. We also analyze the network structure in depth to understand which features drive the decision-making progress and to increase the acceptance and understanding of AI-driven decisions.

Finally, we analyzed USV data from an ongoing phenotypization project and were able to demonstrate the efficacy of the algorithm. Our pipeline improves the overall accuracy of the analysis and provides valuable insights into the distinct ways various deep learning models handle acoustic data.

%Word count: 370

% %detection
% %Deep Squeak; recall 80, precision 97, F1 87 (according to hybrid mouse)
% %Hybrid mouse; recall 98, precision 97, F1 96 (according to hybrid mouse)
% %Pessoa; recall 94, precision 74
% %classification
% %Pessoa; 81.1% accuracy and 80.5% weighted F1
%VocalMat detected over 98% of manually labeled USVs and accurately classified » 86% of the USVs out of 11 USV categories

\section{Material}
\label{sec:Material}

\subsection{Signal acquisition setup}
\label{sec:Signal_acquisition_setup}

Recordings were made using an UltraSoundGate condenser microphone (CM16/CMPA, Avisoft Bioacoustics) connected to a computer via an Avisoft UltraSoundGate USG416H audio device. The USV signals were recorded using Avisoft-RECORDER software (Avisoft Bioacoustics) at a sampling rate of \(250\,\mathrm{kHz} \) and stored in WAV file format. The microphone was positioned close to the ceiling of a \(42x42x42\,\mathrm{cm} \) wide sound-attenuating cube, \(30\,\mathrm{cm} \) above the ground. The room temperature was set to 22°C, and the experimenter ensured that there were no disturbing noises.
%75 words

\subsection{Dataset}
\label{sec:Dataset}

We used USV recordings from an ongoing phenotypization study comparing three heterozygous mouse lines harboring different pathogenic variants in \textit{Nr2f1}. The gene encodes for a transcription factor that plays an important role in neurogenesis.\cite{Bonzano2018-yg} In humans, pathogenic variants in \textit{NR2F1} are associated with intellectual disability, developmental delay, and autism.\cite{Bertacchi2022-zw} Preliminary behavioral data revealed that mice harboring a pathogenic variant in \textit{Nr2f1} resemble the human phenotype, including autism-like behavior, as shown by a strongly reduced number of USVs during pup separation as well as asocial behavior in the three-chamber test (unpublished data). All lines have a C57BL/6J background, as Peleh et al. have shown that the background is suitable for the analysis of USVs in pups.\cite{Peleh2019-ic}

The animals were housed at the Interdisciplinary Neurobehavioral Core (INBC) of Heidelberg University on a 12-hour dark-light cycle. They had ad libitum access to food and water throughout the experiments. Tests were always conducted at the same time.
Recordings were made at three different postnatal (P) ages: P4, P8, and P12. We included both males and females. At P2, animals were tattooed for identification.

Testing of all three lines at the respective timepoints created a dataset comprising a total of 593 recordings (\(5\,\mathrm{min.} \) each), with a total of 160,295 calls (as evaluated by our detection algorithm).

We utilized five distinct datasets in our study:

\textbf{1. 593-Dataset:} This comprises the entirety of 593 recordings and serves as the foundational pool from which the subsequent datasets are derived.

\textbf{2. D-Dataset:} Consists of manually detected calls in a total of 30 recordings and was used for building the detection algorithm.

\textbf{3. VD-Dataset:} Encompasses 6 additional manually detected recordings used for validating the detection algorithm.

\textbf{4. M-Dataset:} Consists of 13 manually detected recordings (from the D-Dataset) that were subsequently manually classified.

\textbf{5. A-Dataset:} Comprises 16 recordings that were automatically detected using the previously established algorithm, followed by manual classification.

Manual detection and classification were conducted using SASLabPro Avisoft (Avisoft Bioacoustics). The detection algorithm development relied on D-Dataset and VD-Dataset, while M- and A-Dataset were used for the classification task. For information on the distribution of call classes in the datasets see \autoref{tab:datasets}.

Upon completing the pipeline, the entire 593-Dataset was analyzed.
%281
%\input{tables/Figure_datasets_overview.tex}
\begin{table*}[tb]
  %Created: 2024-02-15.
  \caption{Call class distribution in the automatically (A) and manually (M) detected dataset.}
  
  \label{tab:datasets}
  \begin{ruledtabular}
  %\resizebox{\textwidth}{!}{
  %\scriptsize
  \begin{tabular}{lccccccc}
    Dataset & size & constant fequency & modulated frequency & frequency step & simultaneous frequencies & short\\

    \hline
    A-Data & \(2013\) & \(261 \ (12.97\%)\) & \(882 \ (43.82\%)\) & \(304 \ (15.10\%)\) & \(165 \ (8.20\%)\) & \(401 \ (19.92\%)\)\\
    M-Data & \(2562\) & \(501 \ (19.56\%)\) & \(760 \ (29.66\%)\) & \(480 \ (18.74\%)\) & \(119 \ (4.64\%)\) & \(702 \ (27.40\%)\)\\

  \end{tabular}
  %}
  \end{ruledtabular}
  
\end{table*}
% #AM p4 5 recordings, p8 5 recordings, p12 6 recordings
% #MM p4 4 recordings, p8 7 recordings, p12 2 recordings

\subsection{Syllable classes}
\label{sec:Syllable_classes}

We adopted the call categories outlined by Scattoni et al. (2008), further refined by Grimsley et al. (2011), and extensively examined by Peleh et al. (2019), as the latter study was also conducted at the INBC.\cite{Scattoni2008-vp,Peleh2019-ic,Grimsley2011-hx} Building on these categories, we consolidated similar classes based on our own observations. This consolidation aimed to augment the number of calls per category, thereby bolstering the reliability of our tests. Specifically, we grouped calls featuring two or more simultaneous frequencies (harmonics and composite), calls exhibiting an absolute modulation of frequency exceeding \(6\,\mathrm{kHz} \) (complex, chevron, upward, and downward), and calls involving a frequency change with no temporal interruption (two-syllable and frequency step). When combined with short calls and those with a constant frequency, this results in a total of five categories:

\begin{itemize}
    \item \textbf{Class 1 Flat:} Calls with a modulation in frequency \textless \(6\,\mathrm{kHz} \).
    \item \textbf{Class 2 Modulated:} Calls with a modulation in frequency \textgreater \(6\,\mathrm{kHz} \).
    \item \textbf{Class 3 Frequency Step:} Calls with an instantaneous frequency change without any interruption in time.
    \item \textbf{Class 4 Composite:} Calls with two harmonically independent components emitted simultaneously.
    \item \textbf{Class 5 Short:} Calls with a duration \textless 5 ms. 
\end{itemize}

\begin{figure*}[ht]
    \begin{center}
    \includegraphics[width = .85\textwidth]{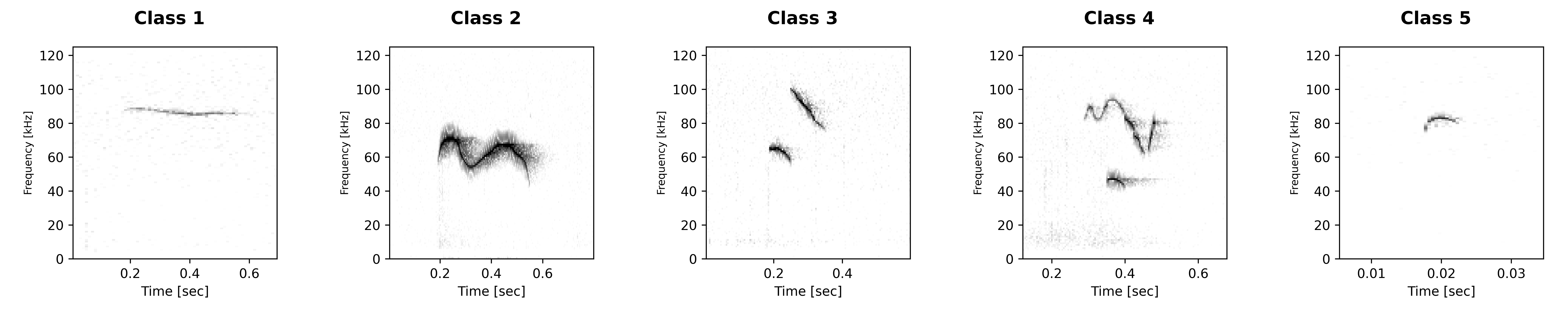}
    \caption{\label{fig:classes}{Overview of the 5 classes.}}
    \end{center}
\end{figure*}

Examples for each class can be seen in \autoref{fig:classes}. Whereas most calls can be easily classified, some are indistinguishable as they fall on the verge between two definitions. Such calls were not included in Datasets M and A. 

%Total words: 192

The pooling was particularly crucial, given that the relative distribution of calls by category emitted by the pups evolves through development and some classes (i.e., two-syllable or chevron) have a significantly lower relative distribution in pups compared to adolescent animals.\cite{Peleh2019-ic} Also, some classes only account for a small fraction of calls. For example, all harmonics, two-syllable, and composite calls together account for less than \(10\,\mathrm{\%} \) of the USVs in C57BL/6J pups.\cite{Scattoni2008-vp} We have pooled call categories that show a similar resemblance to ensure that each category was based on a sufficiently large dataset. Analysis conducted later indicated that pooling did not impede the identification of subtle differences in the quality of USV calls.
%words:98

\section{Methods}
\label{sec:Methods}

\subsection{Structure of the pipeline}
\label{sec:Structure_of_the_pipeline}

\begin{figure*}[ht]
    \begin{center}
    \includegraphics[width = .85\textwidth]{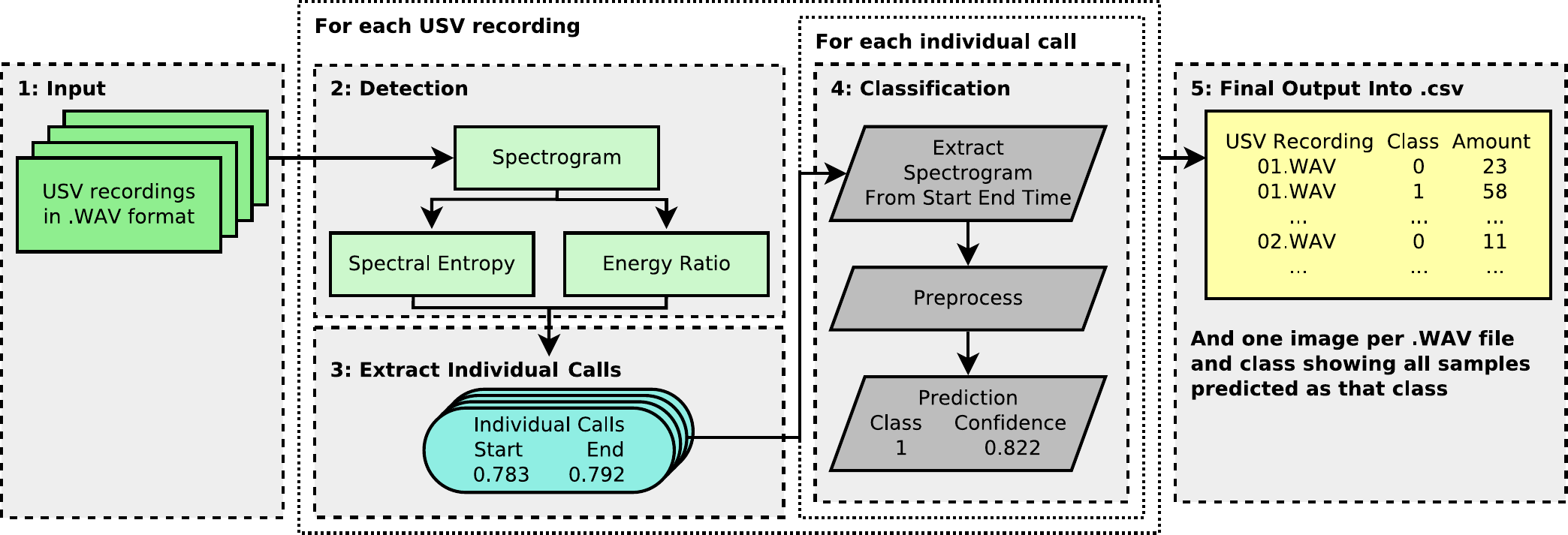}
    \caption{\label{fig:pipeline_overview}{Overview of the final pipeline.}}
    \end{center}
\end{figure*}

Our final goal is to develop a data processing pipeline which allows to load a USV recording and which returns a complete list of calls in this recording along with their classification. Our segmentation and classification code is accessible online through this \href{https://github.com/Nemptis/Neonatal_USV_Detection_Classification.git}{GitHub} repository. \autoref{fig:pipeline_overview} illustrates the comprehensive methodological pipeline.

The two basic building blocks of this pipeline are aimed at 1) extracting short acoustic signals containing only a single call (detection) followed by an analysis block aimed at 2) characterizing the type of call (classification). Finally, these findings are combined in a list of calls along with their classification and returned to the user.

When describing these two blocks in more detail, we have to distinguish between a) the development and b) the application phase of the pipeline. The application phase uses a simple pipeline as depicted in \autoref{fig:pipeline_overview}. In principle, we would like to have software capable of analyzing \(100\,\mathrm{\%} \) of all datasets with \(100\,\mathrm{\%} \) accuracy. However, this is unrealistic in a laboratory setting where different people might conduct experiments, various instruments may be used for recording, and background noise can significantly differ. Therefore, we aim for software that automatically determines a confidence measure for the data. If the confidence is acceptable, we expect reliable classification; if not, the data is flagged for manual inspection. We summarize this approach by defining the \textit{80-90 challenge}.
i.e. we aim at automatically analyzing at least \(80\,\mathrm{\%} \) of all calls with an accuracy of at least \(90\,\mathrm{\%} \). This would result in a reduction of the manual workload by a factor of 5. Furthermore, it allows users to define a minimum level of accuracy tailored to the specific research question. This translates into a threshold that separates calls into those that can be automatically analyzed with the defined accuracy and those that require manual inspection.
%words 219

During the development phase of the software, we also want to analyze and compare different concepts. The concepts used for the detection block are linear in the sense that we use a natural pipeline and develop criteria for optimizing individual steps. For the development of the subsequent classification block, we have primarily used machine learning techniques from deep learning and we have tested several network architectures. For each architecture we need to specify its hyperparmeters, suitable data preprocessing and its output parameters used for evaluating its performance.
%hyperparameters i.e. Learning rate, number of hidden layers, number of neurons in each layer, batch size, dropout rate, etc.
After training of the pipeline, we face the common problem of machine learning, that the structures used by the algorithm for reaching its decision on particular classifications are hidden in the trained network parameters. As an additional feature of our paper, we also aim at visualizing features used by the algorithm in a way that is accessible to human users. 
%109

\subsection{Segmentation, detection}
\label{sec:Segmentation-Detection}

\begin{figure}[ht]
    \begin{center}
    \includegraphics[width = .5\textwidth]{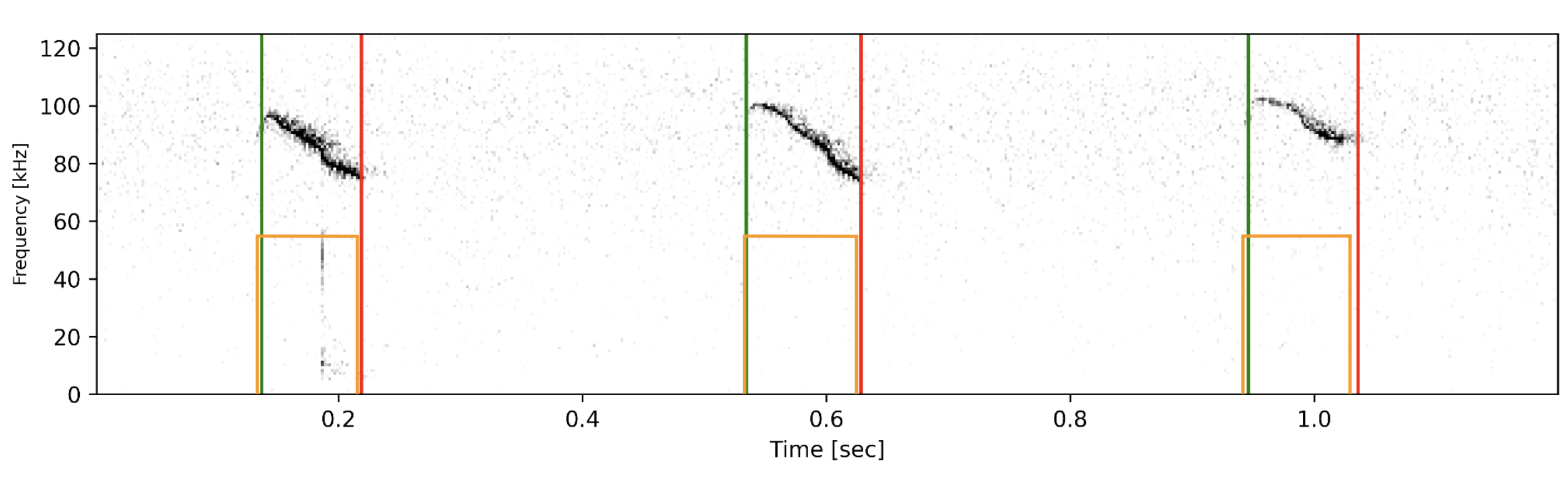}
    \caption{\label{fig:manual_vs_automativ}{A spectrogram displaying three calls, annotated by both the automatic detection (green and red vertical lines) and the manual detection (orange horizontal line).(Color online)}}
    \end{center}
\end{figure}

The first step in our workflow is the detection of individual ultrasonic vocalizations as depicted in \autoref{fig:manual_vs_automativ}. In principle, each recording has to be segmented into periods of vocalization and silence. It is important to note that complete silence is difficult to achieve, and a varying level of background noise cannot be avoided.
%words 42

In order to access discriminating features more easily, the original measurement, i.e., the acoustic pressure wave recorded by the microphone, is converted into a spectrogram. A spectrogram $S(t,f)$ depicts the active frequencies $f$ for each point in time $t$. For a general introduction to spectrograms, one can consult the textbook by Oppenheim on signal processing.\cite{oppenheim_1999}

In our analysis, the spectrogram representation is computed by a short-term Fourier transform (STFT) using a Tukey window with a shape parameter of 0.25, a segment length of 256, and a discrete Fourier transform of length 256 with no overlap. This results in a temporal resolution of slightly more than \(1\,\mathrm{ms} \).
%precisely 1.024
% Tukey window is a mathematical function that is zero-valued outside of some chosen interval
%words 82

Taking the square of the STFT values then yields the energy spectrogram \(S(t,f)\). The detection algorithm relies on the definition of features of \(S(t,f)\) that enable precise determination of the moment a vocalization starts and ends.

Our main feature is spectral entropy. As a measure of the power distribution in the frequency domain, entropy enables differentiation between vocalizations and noise at each time step in the spectrogram. Vocalizations generally only have energy in a very narrow frequency band (low entropy), while noise generally has a dispersed energy spectrum (high entropy).
As background noise is generally restricted to frequencies below \(40\,\mathrm{kHz} \), we compute the spectral entropy in the range between 40 and \(110\,\mathrm{kHz} \) in order to discard low-frequency noise (e.g., mouse movements). We call this restricted part of the spectrogram \(S^h(t,f)\), where \(t\) ranges in discrete time steps between 0 and the length of the recording, and \(f\) ranges in discrete frequency steps between 40 and \(110\,\mathrm{kHz} \). Then the spectral entropy \(H(t)\) is computed by normalization.

    \begin{equation}
    P(t,f) = \frac{S^h(t,f)}{\sum_{f'} S^h(t,f')}
    \end{equation}
    and the subsequent calculation of the standard entropy.
    \begin{equation}
    H(t) = - \sum_f P(t,f) \log P(t,f).
    \end{equation}

To enhance this detection, we incorporated the energy at different frequency ranges as additional features, for example.

$$E^h(t)=\sum_{f=40}^{110}  S(t,f) \ \ \ \mbox{or} \ \ \  E^l(t)=\sum_{f=0}^{39} S(t,f).$$

The first value, $E^h(t)$, denotes the energy of the signal in the typical range of mouse vocalization. The second, low-frequency value, $E^l(t)$, serves as an indicator for the background noise.
However, a simple threshold on these values did not significantly improve the accuracy of the detection. Better results were obtained by thresholding the ratio $R(t)$.
%words 38

% $$E^h(t)=\sum_{40<f<110}  E(t,f) \ \ \ \mbox{or} \ \ \  E^l(t)=\sum_{f=0}^{39} E(t,f).$$

     \begin{equation}
    R(t) = \frac{E^h(t)}{E^l(t)}.
    \end{equation}
%Der Teil zum R(t) hier ist doppelt oder? Wie wäre es wenn man H(t) direkt als zweiten threshold vorstellt? Dann muss man R(t) nur einmal erwähnen.

\begin{figure*}[ht]
    \begin{center}
    \includegraphics[width = .85\textwidth]{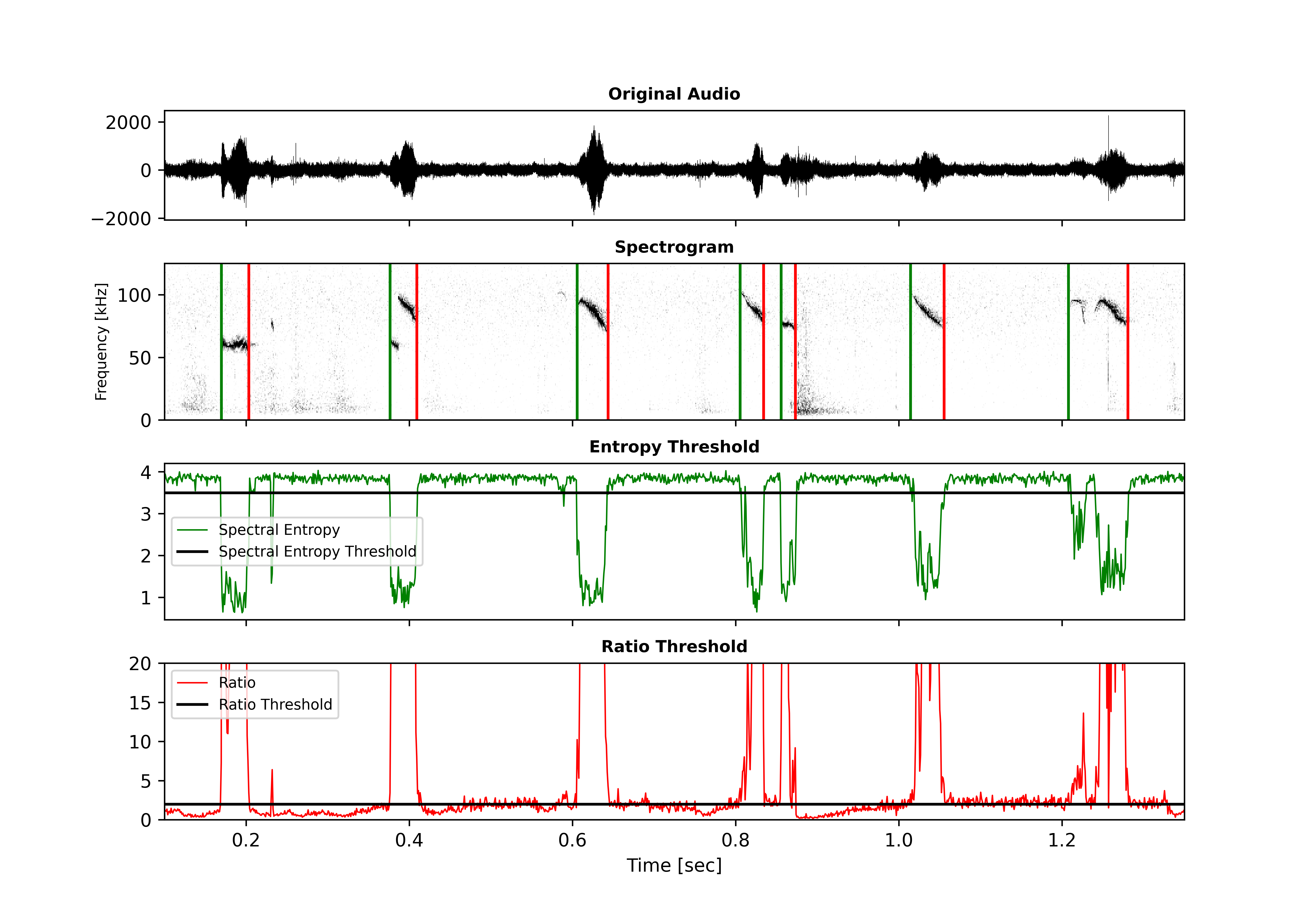}
    \caption{\label{fig:spectrogram_600}{Example of the detection algorithm with corresponding thresholds.(Color online)}}
    \end{center}
\end{figure*}

\autoref{fig:spectrogram_600} illustrates the entropy and the ratio threshold. Combined, they yield an intermediate detection indicator $i(t)$, which takes a value of $0$ or $1$ for each time $t$, indicating whether the spectral entropy as well as the energy ratios are above or below the threshold. In our experiments, we fixed the threshold to 3.5, and a $1$ was assigned to $i(t)$ if $H(t) \leq T_H$ and $R(t) \geq T_R$.

Subsequently, we fuse very short gaps in the detection, which sometimes occur toward the quieter end of a vocalization, and we delete very short detections since they are likely noise. For this purpose, we define two parameters $N_{g}$ and $N_{d}$ and fuse gaps if they are shorter than $N_{g}$ time steps, and delete detections completely if they are shorter than $N_{d}$ time steps. The final output is a list of start and end points of all vocalizations in a recording.

After fixing the parameters of the algorithm, we evaluated its performance on 6 recordings, i.e., \(30\,\mathrm{min.}\) of USV measurements, which were not used in the development of the algorithm. Manual annotation resulted in a set of 2,260 calls. The automatic detection algorithm returned a list of 2,161 calls. 2,146 calls were detected correctly, and 15 calls were false positives. Upon closer inspection, 10 out of those 15 false positives were 'one call as two', i.e., a longer call was detected as two short calls. Moreover, 91 calls which were not detected were merged with another close by call, 'two calls as one'. That is, the calls were detected but not as separate calls. An example of "two calls as one" can be seen in the seventh USV in \autoref{fig:spectrogram_600}. Overall, this results in a recall of \(94.9\,\mathrm{\%} \) and a precision of \(99.3\,\mathrm{\%} \). To the best of our knowledge, this surpasses the quality measures of other existing USV detection schemes as outlined by Pessoa et al..\cite{Pessoa2022-sy}

We also conducted an analysis on how well the algorithm detects the starting and endpoints of the call, which are used for determining the length of the call, but also for extracting the related snippets of the full recording as input for subsequent classification. On the M-Dataset, on average, the starting point detected by the algorithm was delayed by 0.7 ms and the endpoint was also delayed by an average of 5.2 ms. Hence, when extracting the snippets from the full recording for the classification, we added a suitable padding.

Finally, we ran the algorithm on the full set of 593 recordings, which resulted in a set of more than 160,000 calls.

As experimental settings may vary between institutions, we offer an interactive app that allows users to test and optimize the parameters of the detection algorithm individually.
%Total words: 344

\subsection{Neural networks for USV classification}
\label{sec:Neural_networks_for_USV_Classification}

Deep learning concepts based on neural networks have become the gold standard for a wide range of classification tasks.\cite{Aggarwal_2018} In the context of USV classifications, where isolated calls are input, it is reasonable to use feedforward networks.\cite{Sussner_2008} The two main categories of feedforward networks are fully connected networks (FNN, multi-layer perceptron) and convolutional networks (CNN). While FNNs inherently possess higher expressive potential, this comes at the expense of a comparatively high number of parameters that must be optimized during the training phase. CNNs, on the other hand, offer an advantage by sharing a small number of coefficients, i.e., they use the same filter weights when mapping an internal layer to a channel of a subsequent layer. This characteristic facilitates the construction of deeper networks with a restricted number of coefficients, hence making CNNs generally more accessible and stable for optimization. However, it is important to note that CNNs operate under the assumption that discriminating features in the input data are shift and scale invariant, a condition that is only partially met in the context of USV data.

For USV data, we can either pre-compute a set of features and feed the resulting feature values into the network for classification. Typical features based on, for example, average energy levels of the USV in certain frequency ranges, are not shift and scale invariant. Hence, a classification based on pre-manufactured features might benefit from an FNN architecture. If we use the full spectrogram as input, we can assume that the individual types of call may shift and scale somewhat arbitrarily in the time-frequency representation of the spectrogram, and a CNN might be advantageous. A disadvantage of using full spectrograms as input is the comparatively high dimension of the input data. Accordingly, we have tested training both types of network with spectrograms of different levels of downsampling as well as training with pre-manufactured feature vectors.

Our experiments using spectrograms of different sizes proved to be superior when compared with feature-based classification schemes. The only feature, which was explicitly used besides the spectrograms, is the duration of the calls.

It is pertinent to note that each network necessitates data preprocessing. As is customary, we employed networks with fixed input dimensions, while the extracted USV calls exhibited varying lengths. This presents no issue when utilizing pre-manufactured feature vectors, as the challenge of differing USV lengths is transferred to the computation of feature values. However, when employing a full or downsampled spectrogram as input, the spectrograms of USV calls must first be normalized to a standardized dimension.

%Total words: 376

\subsubsection{Fully connected neural networks for USV classification (FNN)}
\label{sec:FNN}

\paragraph{Architecture}
\label{sec:Architecture_FNN}

\begin{figure}[htb]
    \includegraphics[width=0.5\textwidth]{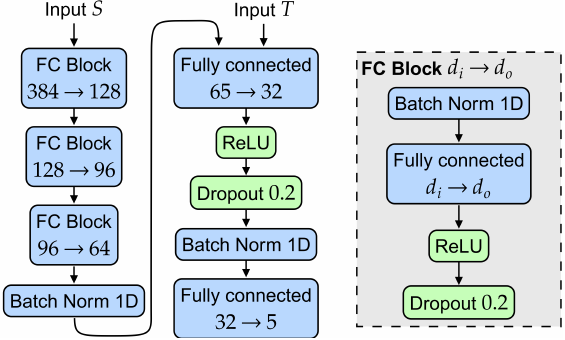}
    \caption{Network architecture of the FNN used for USV classification.}
    \label{fig:FNN_architecture}
\end{figure}

The network is structured around blocks centered on fully connected layers, each consisting of a batch normalization layer, a fully connected layer, a ReLU activation function, and a dropout layer, see \autoref{fig:FNN_architecture}.
Notably, the initial three blocks exclusively receive the downsampled spectrogram \(S\) as input.
However, the relative duration information \(T\) is integrated into the network in the penultimate layer, effectively forming a Y-shaped architecture.
This design artificially increases the importance of the temporal feature.
%words51

\paragraph{Data Preprocessing}
\label{sec:Data_preprocessing_FNN}

Initially, the USV signals are padded with a duration of \(10\,\mathrm{ms} \) at the beginning and end to mitigate any potential signal loss during analysis.
Subsequently, spectrograms are computed from the padded signals using methods outlined in the detection section, resulting in frequency resolutions of \(129\) bins.
The spectrograms are then converted to a decibel (dB) scale, with a clipping threshold set at \(80\,\mathrm{dB} \) to control for signal amplification.
% alles über 80 setzten wir auf 80...
Notably, the horizontal (time) resolution of the resulting spectrograms varies around an average of \(51.4\) pixels, with a standard deviation of \(24.51\) pixels (\(1\) pixel\,\(\approx 1\) ms).
To ensure consistency across samples, the spectrograms are normalized to fall within the range of \([0, 1]\), utilizing standard min-max normalization:
    \[S \gets \frac{S-\min(S)}{\max(S-\min(S))} \ .\]

During training, the spectrograms are shortened by a random number of pixels ranging from \(0\) to \(9\) at the start and end to account for potential inaccuracies in the detected start and end points of the vocalization.
Conversely, during testing, spectrograms are shortened by \(4\) pixels symmetrically, resulting in an effective padding of approximately \(6\,\mathrm{ms} \).
Following this, spectrograms are resized to a constant resolution of \(48\times 8\) pixels.

\begin{figure}[htb]
    \begin{minipage}[t]{0.49\textwidth}
        \includegraphics[width=\textwidth]{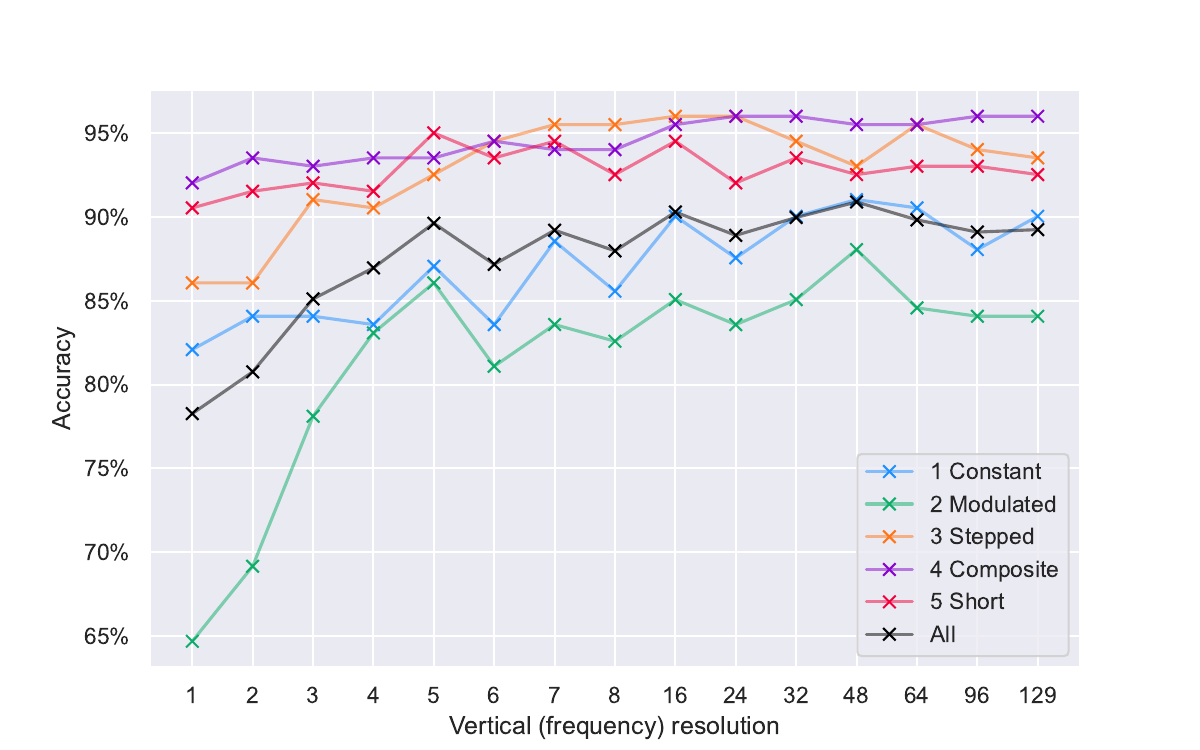}
    \end{minipage}
    \begin{minipage}[t]{0.49\textwidth}
            \includegraphics[width=\textwidth]{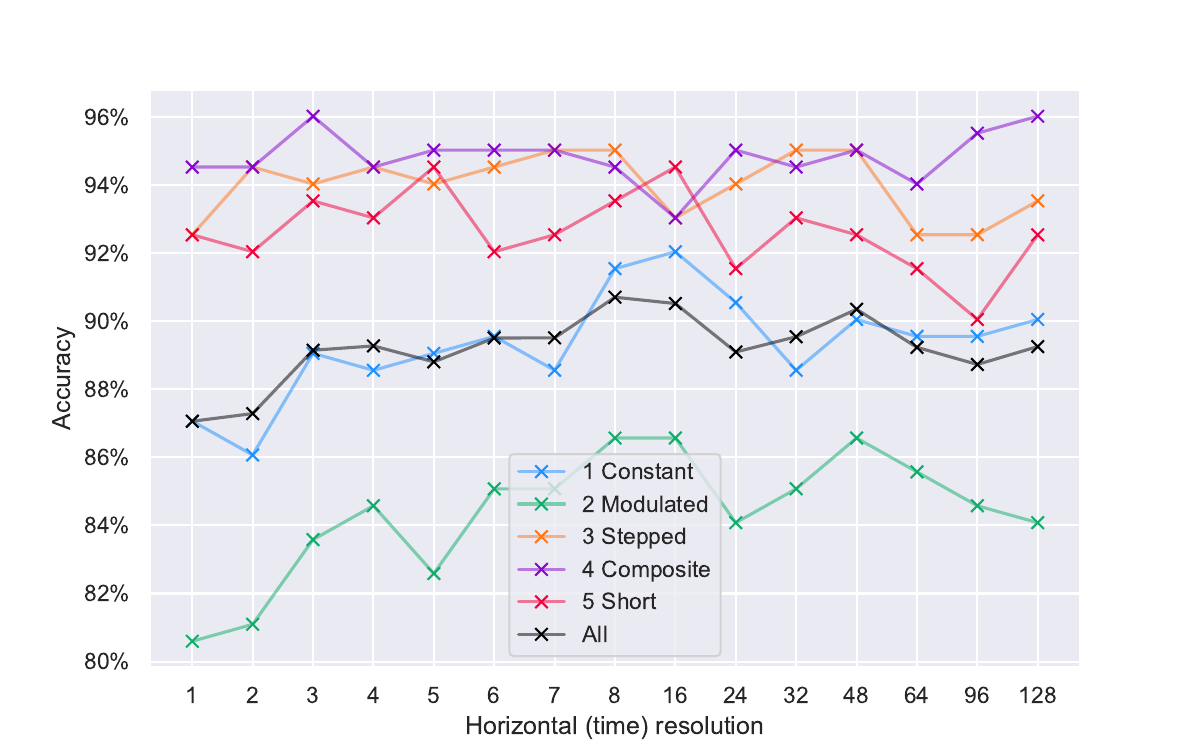}
    \end{minipage}
    \caption{Evaluation of horizontal (time) and vertical (frequency) resolution. For all horizontal resolutions tested, the vertical resolution  was set to \(129\) (the original resolution) and for all vertical resolutions tested, the horizontal resolution was set to \(128\).(Color online)}
\label{fig:Y-Net_res_test}
\end{figure}

This resolution was chosen based on a test cycle for different resolutions. For all horizontal resolutions tested, the original vertical resolution of the full spectrogram was set to \(129\) (the original resolution), and for all vertical resolutions tested, the horizontal resolution was set to \(128\).
Only less than \(4\,\mathrm{\%} \) of our data have an original resolution higher than \(128\) pixels, making this a valid maximum resolution.
Starting from these normalized spectrograms, we then computed down-scaled spectrograms by averaging over neighboring pixels of the spectrogram.
As seen in \autoref{fig:Y-Net_res_test}, the highest accuracy is achieved for a vertical resolution of \(48\) pixels and a horizontal resolution of \(8\).
This was flattened into the 1-dimensional feature vector of length \(384\) and used as the input \(S\).

To normalize sequence lengths, each USV duration is divided by the maximum observed length, which is approximately \(150\,\mathrm{ms} \), resulting in the relative duration \(T\).
As usual, we randomly add noise to the training data for stabilizing the learning process; we used \(5\,\mathrm{\%} \) additive noise with a normal distribution.

%words: 342

%da input t bereits normalisiert ist muss T nicht mehr durch batch norm geschleust werden...

\paragraph{Training and Regularization}
\label{sec:Training_and_REgularization_FNN}

For training, we use regularization through the use of the Adam optimizer with weight decay, as proposed by Loshchilov and Hutter, with a specific learning rate set at \(0.0001\).\cite{loshchilov2019decoupled}
To prevent overfitting and promote generalization, a dropout rate of
\(20\,\mathrm{\%} \) is applied after each Rectified Linear Unit (ReLU) activation function.
The neural network has \(71600\) trainable parameters.
%words 34

%%%Johann Notes:
%Dropout is a regularization technique used in neural networks during training. It involves randomly "dropping out" (i.e., setting to zero) a fraction of the units in a layer at each update during training. This helps prevent overfitting
%Learning rate is a hyperparameter that determines the step size at each iteration while moving toward a minimum of the loss function. It controls how much the model's weights are updated during training. Too high a learning rate may cause the model to converge too quickly, potentially skipping the minimum, while too low a learning rate may result in slow convergence or getting stuck in a local minimum.
%Weight decay is a regularization term added to the loss function during training to penalize large weights. It helps prevent overfitting by discouraging the model from fitting the training data too closely. The weight decay parameter controls the strength of this penalty. It is essentially a form of L2 regularization where the sum of squared weights is added to the loss function.

\subsubsection{CNN and ViT}
\label{sec:CNN_and_ViT}

In this section we describe the CNN and vision transformer (ViT) architectures, the data preprocessing and the regularization we use for their training.
%words 17

\paragraph{Architectures}
\label{sec:Architecture_CNN}

The classification experiments using convolutional networks were done with three classical off-the-shelve network architectures (ResNet34 \cite{resnets}, ResNet50 \cite{resnets} and EfficientNet-B5 \cite{efficientnetv1}) and a customized architecture based on ResNet blocks.
Additionally we used a model with a ViT architecture, ViT-B/16.\cite{ViT}

\begin{figure}[tb]
    \includegraphics[width=0.5\textwidth]{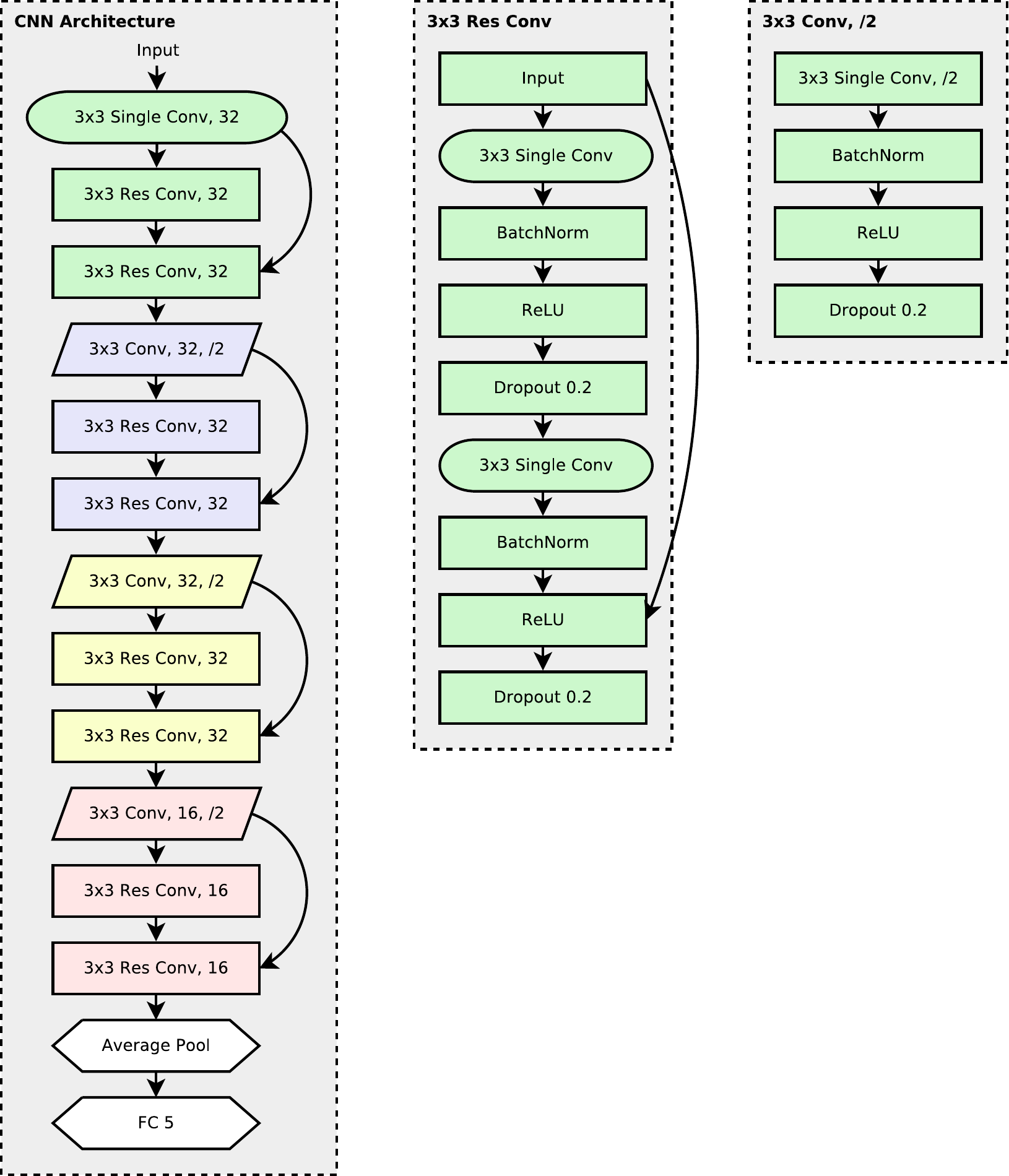}
    \caption{\label{fig:model_architecture_custom_cnn}{Model architecture of the custom CNN.}}
\end{figure}

As inspiration for our own customized CNN, we took the paper by He et al., where such an architecture was developed for tasks in computer vision.\cite{resnets} As can be seen in \autoref{tab:model_params}, our custom CNN is much smaller than off-the-shelve architectures, which is preferable in terms of trainability and interpretability. The specific architecture of the customized CNN can be seen in \autoref{fig:model_architecture_custom_cnn}.

\begin{table*}[tb]
    \caption{\label{tab:model_params} Number of trainable parameters for the different models.}
    \begin{ruledtabular}
    \begin{tabular}{l|cccccc}
    Model & FNN & Custom CNN & ResNet34 & ResNet50 & EfficientNet-B5 & ViT-B/16 \\
    \hline
    Trainable Parameters & \num{71600} & \num{149354} & \num{21800242} & \num{25567282} & \num{30400034} & \num{85728773} \\
    \end{tabular}
    \end{ruledtabular}
\end{table*}

For details on the hardware and software refer to the respective section in the appendix.
%words 84

\paragraph{Data Preprocessing}
\label{sec:Data_preprocessing_CNN}

We adjust the data preprocessing pipeline compared to the one for the FNN model.
Since CNNs can handle larger input sizes easier than a fully connected neural network (due to shared parameters), we opt to use the spectrograms at full resolution.
Due to batching the data in training, all the spectrograms need to have the same resolution at the end of the pipeline, which we achieve via cropping too long and padding too short spectrograms to the same size. We avoid resizing which would probably make validation more difficult due to the additional distribution shift.

\begin{figure*}[tb]
    \includegraphics[width=0.85\textwidth]{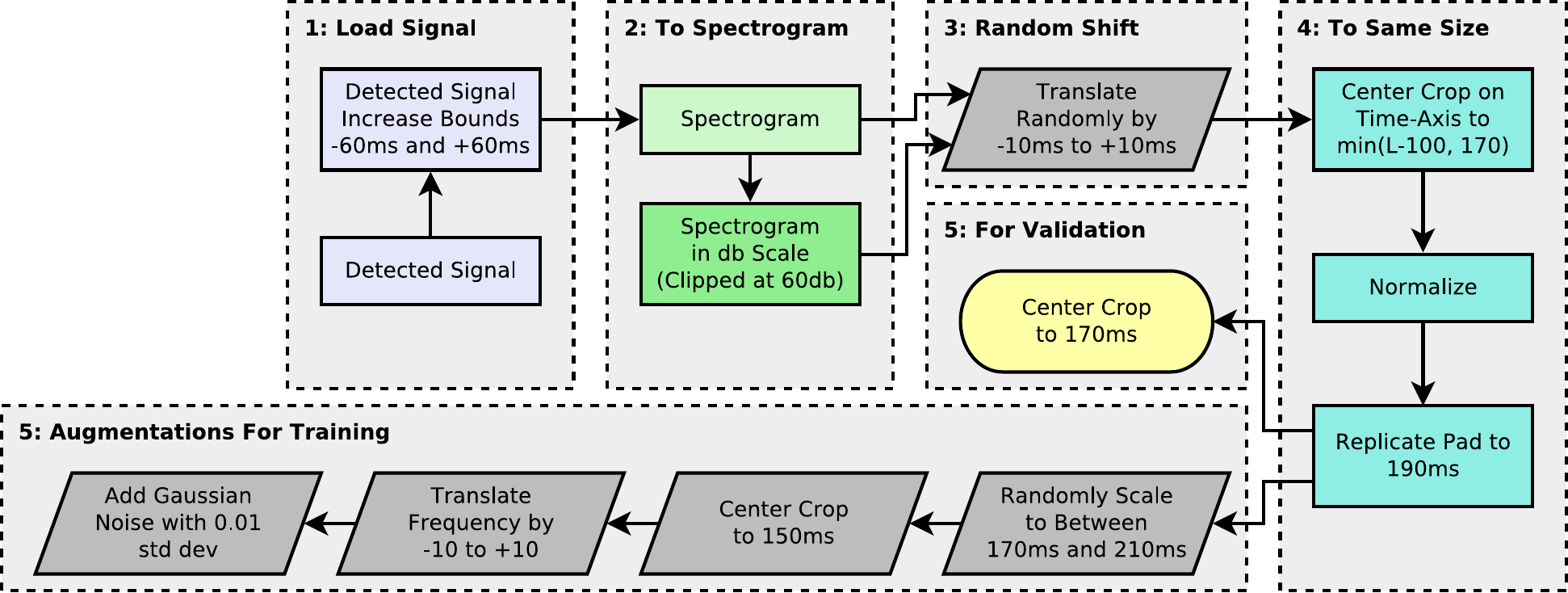}
    \caption{\label{fig:data_loading_custom_cnn}{Data loading pipeline for the CNNs and ViT.}}
\end{figure*}

Our data loading and preprocessing pipeline is shown in Figure \ref{fig:data_loading_custom_cnn}.

The five grey boxes describe transformations only done in the training pipeline (augmentations are only used in training), and the yellow box is only executed in validation phase.

The key features of this preprocessing are as follow. In the first step, we increase the bounds of the signal we obtained from the automatic detection by 60ms on both sides, extending the signal duration before and after the detected signal (this is to compensate for the random shift).
Then we transform the signal into a 2d spectrogram, and use both the original spectrogram and a spectrogram rescaled to a decibel scale (while clipping its lower end so that the whole spectrogram has a \(60\,\mathrm{dB} \) range). Both spectrograms are concatenated, i.e. merged into a single data block, and used as input for the network.
In order to improve robustness and to avoid overfitting, we then randomly translate the spectograms by up to 10ms on the time axis as augmentation in the training cycle (we do not want the model to overfit on the position of the call, since the detection boundaries can vary).

As previously mentioned, due to batching the data in training all spectrograms need to be transformed into the same shape.
We do this by determining the mid-point in time of the spectrograms and by extracting the central part of the spectrograms, i.e. the data is center cropped to min ($\ell  -100$, $170$) with $\ell$ being the duration of the spectrograms in milliseconds (crop too long spectrograms to a maximum size of 170).
Then we subtract the mean and divide by the standard deviation of the labeled dataset to normalize the data.
After that we use replication padding to pad the spectrograms to a length of \(190\,\mathrm{ms} \), at this point all spectrograms have the same shape of $2 \times 201 \times 190$, as required for batching.

For the validation of the pipeline, the spectrograms are only center cropped to a length of 170, thus completing the pipeline.
For the training pipeline, we use additional data augmentations, to improve robustness and to avoid overfitting.
We rescale the spectrograms on the time axis randomly between \(170\,\mathrm{ms} \) and \(220\,\mathrm{ms} \), center crop them to \(150\,\mathrm{ms} \), randomly translate them by -10 to 10 on the frequency axis and finally add gaussian noise with a standard deviation of $0.01$.

Before we output the spectrograms, we stack the time feature into them as an additional channel (replicate the time feature value over all spatial positions), resulting in a final spectrogram shape of $3 \times 201 \times 170$ for validation, respectively $3 \times 201 \times 150$ for training.
%words 399
%

\paragraph{Training and Regularization}
\label{sec:Training_and_regularization_CNN}

For regularization, we use Adam with weight decay with the default learning rate of 0.001 as optimizer.\cite{loshchilov2019decoupled}
Further we use label smoothing in the loss (we use 0.05 and 0.9 as targets instead of 0.0 and 1.0).
For the custom CNN, we additionally use twenty percent dropout after each activation function. Meaning that for each element there is a \(20\,\mathrm{\%} \) chance that the value is set to zero.\cite{JMLR:v15:srivastava14a} 
For the other models, we use their default settings, which means no dropout in ResNet34, ResNet50 and ViT and stochastic dropout of resnet blocks in EfficientNet-B5 with a maximum probability of 0.2 (which gets linearly scaled by the layer of the model, so for a given layer the dropout probability would be $0.2 \cdot \frac{\text{layer id}}{\text{number of layers}}$).
The stochastic dropout in the EfficientNet-B5 model works for each row of the batch, i.e. an entire input is zeroed out.
%words: 361

\section{Results}
\label{sec:Results}

For the presentation of our results we distinguish between the fully automated and the semi-automated case. The fully automated setting refers to the scenario where a set of USV recordings is provided as input, and a list of individual calls along with their classifications is returned. This case is presented in the next subsection. On the other hand, the semi-automated case aims to reduce the manual workload and to achieve a desired recall, typically higher than what is achievable with the fully automated setting. In the semi-automated setting, we require an additional measure that evaluates the quality of the classification result. Depending on this quality measure, the system either returns the classification result or retains the particular call for further manual inspection.

In the following subsection, we focus on the results of the classification step. The results of the detection step were already presented in \autoref{sec:Segmentation-Detection}.
%words 129

\subsection{Evaluation metrics for classification}
\label{sec:Evaluation_metrics_for_classification}

The quality of the different neural network architectures for the classification of neonatal USV calls are based on 10-fold cross-validation with A-data set plus using the whole M-data set for training.
The A-data set is randomly split into 10 disjunct splits (all neural network architectures use the same 10 splits) with the union of those 10 splits resulting in the whole A-data set. Then we train 10 models (per network architecture), each time using 9 splits for training and 1 split for validation, where we choose a different validation split for each of the 10 models. The reported quality measures are averaged over those 10 models.

The usual quality measures (accuracy, recall, precision, specificity, F1-score) are computed for either class-wise classification or for determining an overall score.
For class-wise classification we treat our multi-class problem as a binary-class problem, we determine the quality measures in a one vs all manner, i.e. for each class we treat all the other classes as the same class.
E.g. true positive refers to the number of calls, that were correctly assigned to this class, true negatives refer to all calls not in this class which were assigned to any of the other classes, similar for false positives and false negatives. We then use the standard definitions:
%words 121
\begin{align*}
    \text{Accuracy} &= \frac{\text{True Positives} + \text{True Negatives}}{\text{All}}\\
    \text{Recall} &= \frac{\text{True Positives}}{\text{True Positives + False Negatives}}\\
    \text{Specificity} &= \frac{\text{True Negatives}}{\text{True Negatives + False Positives}}\\
    \text{Precision} &= \frac{\text{True Positives}}{\text{True Positives + False Positives}}\\
    \text{F1-score} &= 2 \cdot \frac{\text{Precision} \cdot \text{Recall}}{\text{Precision + Recall}}
\end{align*}
The recall focuses on capturing all actual positive instances, while the specificity focuses on correctly identifying all actual negative instances.
The precision penalizes false positives and the F1-score is a measure for the balance of precision and recall.
%words 28

We compute the overall accuracy as the ratio between correct predictions and all predictions, as previously described.\cite{Pessoa2022-sy, 10.7554/eLife.59161}
For the other values, we compute them per class in a one vs all manner, and report the weighted mean over the classes, as described in Pessoa et al..\cite{Pessoa2022-sy}
In a multi-class setting, the class weighted binary recall results in the ratio of correct predictions to all predictions, i.e. for multi-class our reported recall is the same as the reported accuracy.
%
%In this paper, we choose to refer to the ratio of correct predictions to all predictions as recall (which is usually referred to as accuracy).\cite{Pessoa2022-sy, 10.7554/eLife.59161}
%words 76
%
% Not sure how to write this best, in all our tables we do not report the results of the metrics directly, but first multiply them by 100. E. g. 77.35 Recall would actually be 0.7735
The results are given in percent.

\subsection{Fully automated classification}
\label{sec:Fully_automated_classification}

In the following section, we will compare the performance of the FNN, the Custom CNN (with different variants of preprocessing and downsampeling), the ResNet34, ResNet50, EfficiencyNet-B5 and the ViT. The mentioned networks differ in their architecture and also in their size, see Table \ref{tab:model_params}).
A (*) next to the model name indicates that the smooth spectrograms were normalized individually in the data preprocessing pipeline (i.e. the mean of each smooth spectrogram gets set to 0 and the standard deviation to 1), as opposed to normalizing them over the global mean and standard deviation of all the smooth spectrograms in the M-data set.
%%%TODO: brauchen wir hier die Erklärung zu dem * wirklich, bzw können wir da nicht einfach weglassen?

 As already mentioned, we use tenfold cross-validation on the A-data set for evaluation. Hence, for each network architecture, we trained ten models, and evaluated them separately.
 For class-wise evaluation for each of the these models we compute accuracy, precision, recall, specificity and F1-score all binary in a one-vs-all approach.
 The global numbers reported for precision, recall, specificity and F1-score are then the weighted sums over all five classes, i.e. we weight the quality measures for each class by the number of samples in this class and sum them up. These values are averaged over all ten runs of the cross-validation, which also allows us to compute the variance over these ten runs.
 The global recall is the ratio of correct predictions to all predictions.
The results are displayed in \autoref{tab:classification_results}.  
We note a few observations before going into more detail on the performance of the custom CNN.
The EfficientNet-B5 model performs best, but with more than 30 million trainable parameters it is also the largest of the convolutional models we used.
The vision transformer (ViT-B/16) does not perform well, which is expected due to the low amount of data we have for training (vision transformer generally require larger amounts of training data compared to CNNs, in the original paper they were pretrained on a dataset consisting of 300 million images, whereas we only have 4374 spectrograms for training). \cite{ViT}
For ViT-B/16, we rescaled the spectrograms to a final size of 160x160, as the image size needs to be a multiple of the patch size (16).
The custom CNN performed similar to the EfficientNet-B5 with only 149,000 trainable parameters, therefore for the following interpretability section we focus on the custom CNN model.

For the custom CNN we add the results for some variants.
E.g. the use of data augmentations and regularization in the model, gives us an increase of \(3.92\,\mathrm{\%} \) in recall (in \ref{tab:classification_results} 'Custom CNN no aug reg' uses no augmentation or regularization and has \(3.92\,\mathrm{\%} \) lower recall than the original model). Additionally, downsampling the spectrogram to $25 \times 8$ drastically reduces performance ('Custom CNN 25 $\times$ 8'), hence we decided to work with full spectrograms for classification with all CNNs. Not clipping dB values in the spectrograms ('Custom CNN no dB limit') did slightly increase performance measures but at the expense of a higher variance, therefore we included dB clipping. Also, duplicate channels did not yield a better performance ('Custom CNN $\times$ 2 channels').
Hence, we fixed the custom CNN network as described in the previous section. The performance is presented in \autoref{tab:classification_results}. For a class-wise evalation of the cutom CNN see \autoref{tab:classification_results_per_class}. Detailed information on the FNN can be seen in \autoref{tab:classes-metrics}.
%words 436

\begin{table*}[tb]
\caption{\label{tab:classification_results} Results of the classification models, evaluated on the 10 fold cross validation test data.}
%%%accuracy hier richtig berechnet Meeting2

%\begin{ruledtabular}
%\resizebox{\textwidth}{!}{%
\begin{tabular}{lccccc}
\hline
\hline
    Model & Precision & Specificity & F1 Score & Accuracy \\
\hline
FNN & $78.14 \pm 2.48$ & $90.49 \pm 1.14$ & $77.14 \pm 2.61$ & $ 77.35 \pm 2.55$ \\
Custom CNN & $87.02 \pm 1.46$ & $94.68 \pm 1.21$ & $86.69 \pm 1.46$ & $86.79 \pm 1.45$ \\
Custom CNN no aug reg & $83.28 \pm 1.62$ & $93.09 \pm 1.79$ & $82.37 \pm 2.00$ & $82.86 \pm 1.75$ \\
Custom CNN no db limit & $87.24 \pm 2.71$ & $94.80 \pm 1.40$ & $67.76 \pm 2.80$ & $86.88 \pm 2.71$ \\
Custom CNN 25x8 & $70.03 \pm 2.64$ & $85.60 \pm 1.87$ & $68.75 \pm 2.34$ & $70.09 \pm 1.94$ \\
Custom CNN x2 channels & $86.81 \pm 1.62$ & $94.69 \pm 0.92$ & $86.41 \pm 1.69$ & $86.49 \pm 1.61$ \\
ResNet34 (*) & $86.75 \pm 1.24$ & $94.64 \pm 1.10$ & $86.21 \pm 1.42$ & $86.34 \pm 1.39$ \\
ResNet50 (*) & $87.27 \pm 1.51$ & $94.86 \pm 0.11$ & $86.80 \pm 1.43$ & $86.88 \pm 1.55$ \\
EfficientNet B5 (*) & $\textbf{87.58} \pm 1.43$ & $\textbf{95.27} \pm 0.07$ & $\textbf{87.21} \pm 1.35$ & $\textbf{87.28} \pm 1.36$ \\
Custom CNN (*) & $86.10 \pm 1.46$ & $94.34 \pm 1.00$ & $85.67 \pm 1.33$ & $85.84 \pm 1.35$ \\
ViT-B/16 (*) & $57.75 \pm 4.30$ & $76.45 \pm 2.89$ & $55.84 \pm 3.20$ & $61.89 \pm 2.38$ \\
\hline
\hline
\end{tabular}
%}
%\end{ruledtabular}
\end{table*}

% (*) behind the model means that here the smooth spectrograms are normalized individually, to mean 0 and standard deviation 1 (this was necessary for the other cnn models (ResNet50...) to work). A problem with the smooth spectrograms is that their scale is very different (by a factor of around 1000, which make them kind of unusable for the networks and seem to lead to trouble in the beginning of the training -> the custom cnn has eventually learned to not use them).
% The ViT-B/16 model evaluates at a time size of 150 (instead of 170) and the spectrograms are resized to (160x160) to be a multiple of the patch size 16.
% The ViT-B/16 gets the time feature only as third channel as input, for the ResNet34, ResNet50 and EfficentNet B5 the time feature is used as input both as third channel in the spectrogram, and as additional channel in the input of the final linear classifier at the end of the model.
% Currently done ignoring the class weights. So computes the stats for each of the 10 models per class -> tensor of shape (10, 5) per stat, then computes mean or std over the 0 dim (the 10 models) and reports that.
% the val data can have different number of samples per class i.e. for one model there could be 5 samples of class x whereas for another model there are 50 samples, but both count the same in the metrics here, it is not weighted with the amount of samples
% -> not really reducing standard deviation when using weighting

%Custom CNN evaluation per class
\begin{table}[tb]
    \caption{\label{tab:classification_results_per_class} Results of the custom CNN model shown per class, evaluated on the 10 fold cross validation test data. All the metrics (including Accuracy) in the Table are computed in a one vs all manner.}

    %\begin{ruledtabular}
    \resizebox{0.5\textwidth}{!}{%
    \begin{tabular}{lccccc}
    \hline
    \hline
     Class & Precision & Recall & Specificity & F1 Score & Accuracy  \\
    \hline
    1 & $78.07 \pm 5.68$ & $80.14 \pm 9.93$ & $96.50 \pm 1.41$ & $78.77 \pm 5.63$ & $94.39 \pm 1.75$ \\
    2 & $89.41 \pm 3.13$ & $91.68 \pm 3.49$ & $91.36 \pm 3.00$ & $90.45 \pm 1.61$ & $91.55 \pm 1.37$ \\
    3 & $85.17 \pm 5.84$ & $79.73 \pm 7.82$ & $97.47 \pm 1.32$ & $82.14 \pm 5.38$ & $94.83 \pm 1.65$ \\
    4 & $86.05 \pm 9.58$ & $75.56 \pm 8.09$ & $98.75 \pm 1.19$ & $80.00 \pm 6.23$ & $96.92 \pm 1.19$ \\
    5 & $89.39 \pm 3.25$ & $89.51 \pm 5.01$ & $97.39 \pm 0.78$ & $89.37 \pm 3.10$ & $95.88 \pm 0.97$ \\
    \hline
    \hline
    \end{tabular}
    }
    %\end{ruledtabular}
\end{table}

\subsection{Semi-automated USV analysis}
\label{sec:Semi_automated_USV_analysis}

Alternative to the fully-automated approach one can use the last layer of the network, which gives a pseudo-probability for each of the five classes, as a quality indicator to sort out calls where the algorithm is not sure. Allowing the user to define a threshold $p \in [ 0,1 ]$ and accept the algorithms finding only, if the algorithmic  pseudo-probability exceeds $p$ for at least one class.
For each value of $p$, we determine the number of calls reaching pseudo-probabilities above the threshold and report the resulting recall.

\begin{figure*}[ht]
    \includegraphics[width=1.0\textwidth]{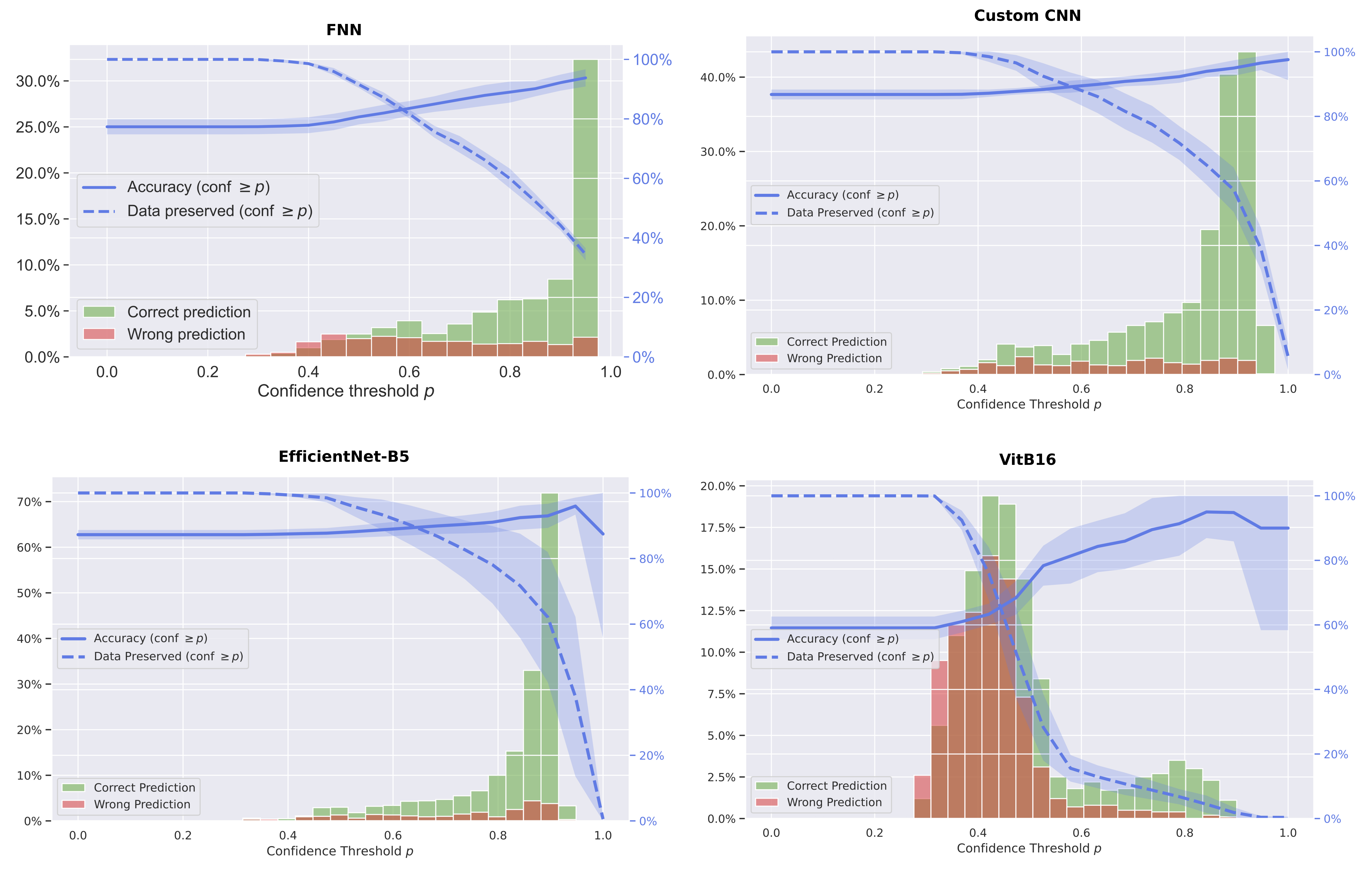}
    \caption{\label{fig:graph_prob}{Mean and standard deviation of the recall on the validation data of the 10-fold cross validation sets while ignoring the predictions with the lowest confidence.(Color online)}}
\end{figure*}

We have plotted the resulting curves in \autoref{fig:graph_prob}. I.e. for the EfficientNet-B5 model a threshold of $p = 0.8$ allows to classify \(71.7\,\mathrm{\%} \) of the calls with an recall of \(92.5\,\mathrm{\%} \). As described, we were aiming for the $80-90$ challenge, i.e. we aspired to classify \(80\,\mathrm{\%} \) of all calls automatically with a recall of at least \(90\,\mathrm{\%} \).
%words 210

To further examine the effects of limiting the evaluation to samples, for which the neural network outputs a relatively high confidence, we developed the graphic shown in \autoref{fig:graph_prob}.
First, let us focus on the histograms, shown in red and green with their values represented on the left-hand y-axis.
%%%Ja, aber schwierig formuliert
For a total of 21 equally spaced confidence intervals, we calculate the relative amount of samples, with a prediction confidence falling into the given interval.
This is done for correctly and wrongly classified samples separately.
It is evident that most correctly classified samples have a high confidence and vice versa for the wrongly classified calls.\par
This motivates a confidence dependent evaluation of the neural network, which represents a generalization of the before mentioned $80-90$ challenge. 
To visualize this, we plot the recall (solid blue line) and the amount of data preserved (dashed blue line) in dependence of the pseudo-probability $p$.
%%%TODO: Muss im daigram nochmal umbenannt werden 
For example, if all samples with a prediction confidence of less than \(0.6\) are omitted, \(81.7\,\mathrm{\%} \) of the data is left, for which the network reaches a recall of \(83.5\,\mathrm{\%} \).

In summary, the proposed data analysis pipeline is capable of selecting and classifying \(82.7\,\mathrm{\%} \) of all calls with a recall of \(90.4\,\mathrm{\%} \) (with a threshold of $p=0.7$). The remaining calls are separated and can be analyzed manually. In total, this allows an almost arbitrarily high recall while still saving a substantial time of manual labor.

\subsection{Interpretability of the Custom CNN}
\label{sec:Interpretability_of_the_custum_CNN}

In this section, we aim to analyze how the classification network reaches its decisions. We start by displaying channel visualizations of the different layers, followed by saliency maps that highlight which parts of the input the model relies on for its prediction.

\subsubsection{Channel Visualization}

Channel visualizations indicate which structures are analyzed at different channels. These visualizations serve as basic building blocks, and the output values of these channels can be interpreted as a measure of how strongly the input correlates with these structures. A description of how the visualizations are computed can be found in the appendix. For example, Figure \ref{fig:channel_resnet2} displays several channels that are tuned for analyzing constant, increasing, or decreasing frequency content. The channel at the bottom left seems to capture signals with a discontinuity in frequency. The multiplicity of similar channels might indicate that similar results could be achieved with a smaller network. However, these similar structures are combined with different follow-up structures in the subsequent layers, so this multiplicity is necessary for allowing a subtle analysis.

\begin{figure}[tb]
    \includegraphics[width=0.5\textwidth]{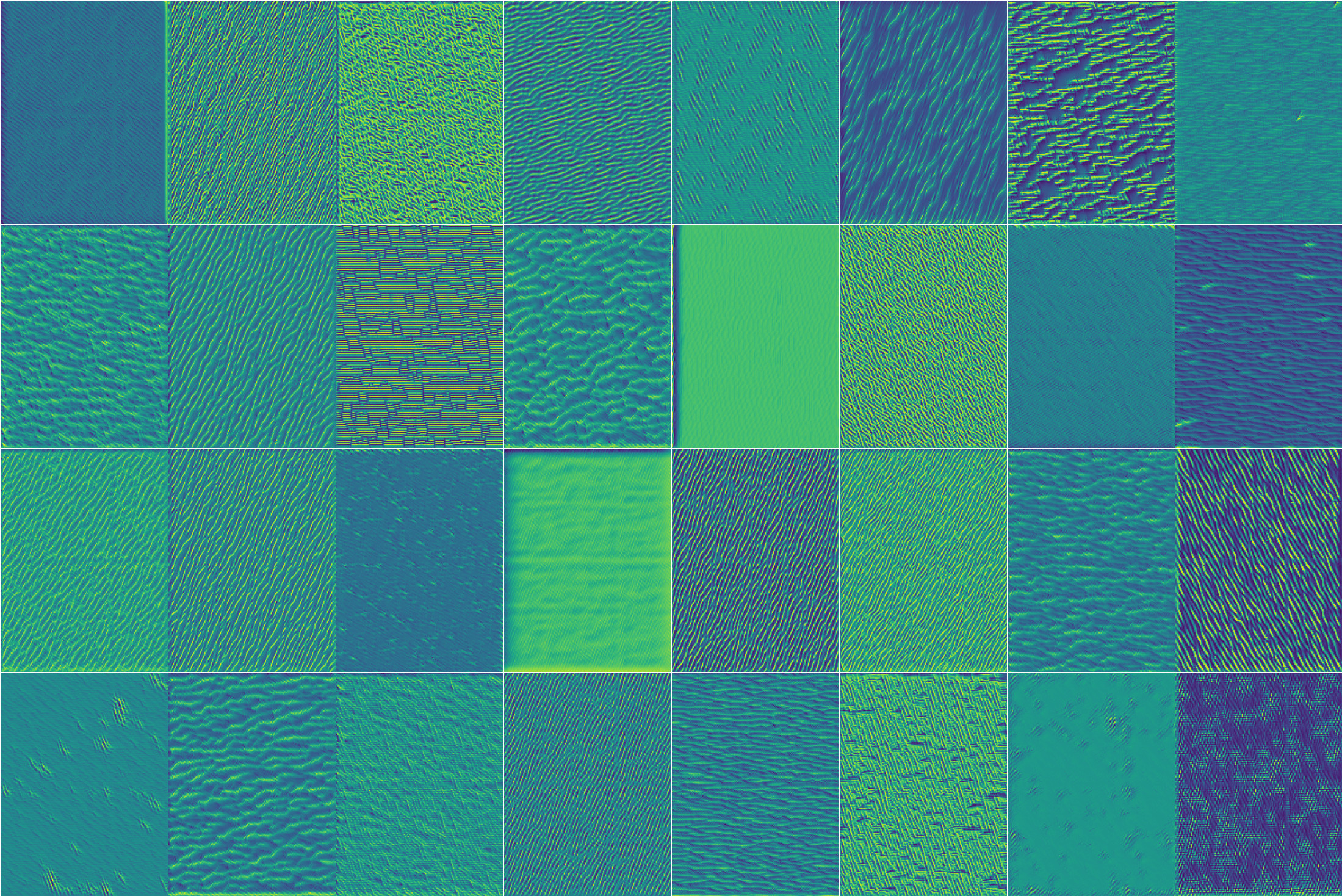}
    \caption{\label{fig:channel_resnet2}{Visualisation of the activations in the 32 channels of the first layer of the network.}}
\end{figure}

The structures in the channel visualizations of the later layers become increasingly difficult to analyze. The channel visualizations for all layers of the network are depicted in \autoref{sec:Appendixes}.
%words 130

\subsubsection{Saliency Maps}
\label{sec:Saliency_Maps}

As a second approach for explaining how the network functions, we used saliency maps which are computed as follows. For a given input one determines how strongly every pixel contributes towards the final classification.

As method to generate the saliency map, we use Integrated Gradients \cite{pmlr-v70-sundararajan17a} combined with SmoothGrad \cite{SmoothGrad}. For details, refer to the appendix. 

\begin{figure}[tb]
    \includegraphics[width=0.5\textwidth]{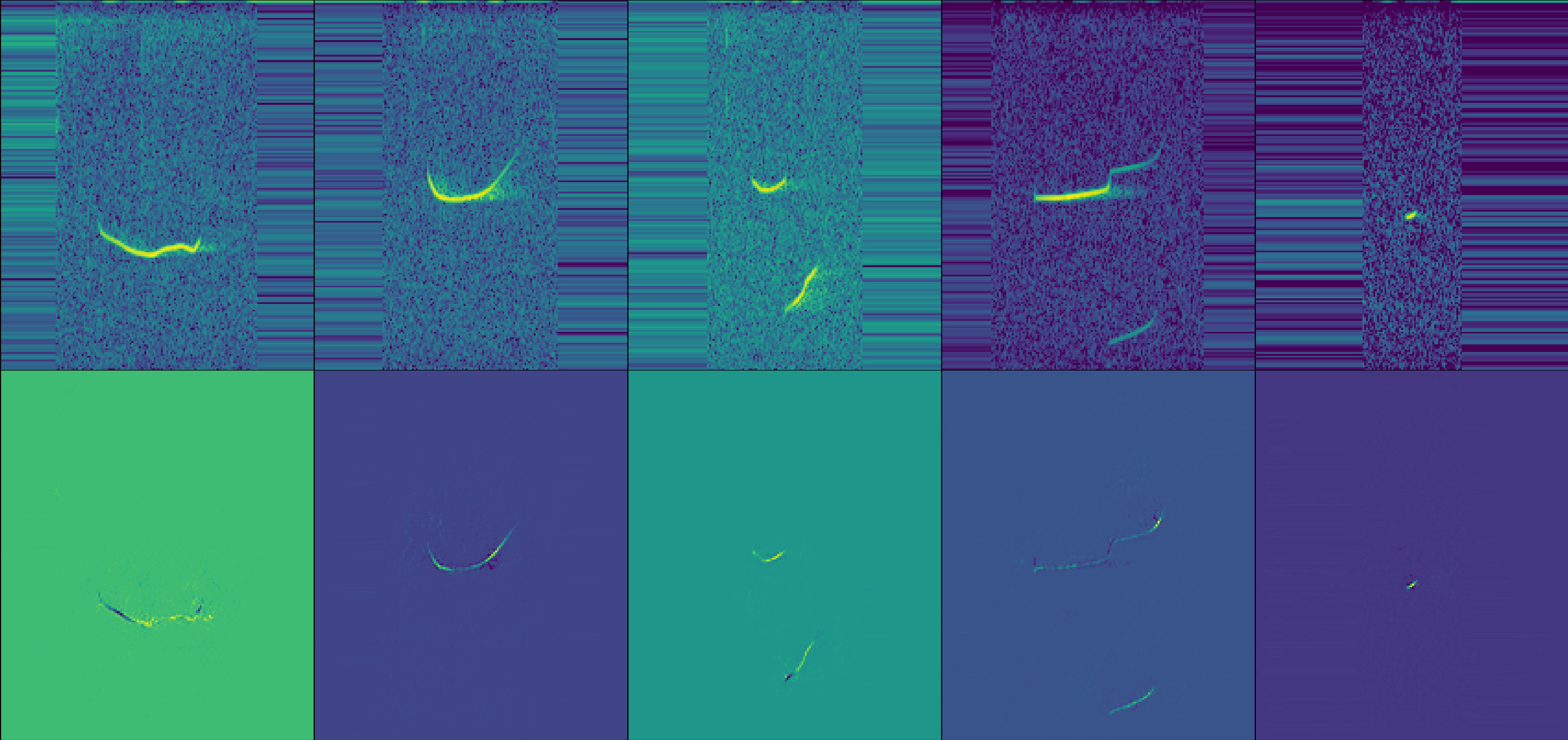}
    \caption{\label{fig:saliency_map}{Top row: input data for calls from different classes, the effect of the padding structure for normalizing the inputs to fixed dimensions are visible; bottom row: The resulting saliency maps show that the network indeed is looking at core structures of the signals and neglects the noise background. The calls classes are: constant, modulated, frequency step, composite and short call in that respective order.}}
\end{figure}

As examples, we plot the saliency maps for prototypical vocalizations for every class, see \autoref{fig:saliency_map}. As expected, a well trained network looks at the prototypical shapes of the signal and focuses on the main frequencies at each time instant. It neglects e.g. the blurring in frequency and the background structures almost completely. We take this as a confirmation that the network is well trained for the task at hand.
%words 97

\section{Discussion}
\label{sec:Discussion}

In this paper, we have developed two neural network architectures (FNN, custom CNN) for USV classification and we compered them with much larger off-the-shelf network architectures (several CNNs, Visual Transformer ViT).  The best results in terms of overall accuracy were achieved by the 'EfficientNet B5' architecture. However, ResNet50 as well as our custom CNN achieve a comparable accuracy which is within the range of \(0.5\,\mathrm{\%} \) of EfficientNet B5. 

Alternative architectures, such as ViT and FNN, did not achieve comparable performance levels. The reasons for these inferior results vary. ViT, a large and powerful and architecture, demands vast quantities of training data. In the present setting, overfitting to the relatively limited training data is nearly inevitable, hindering further improvements in training. As for FNN, we posit that the inherent but constrained translation invariance of the classification task, wherein slight shifts in signal frequency or timing do not alter classification outcomes, provides convolutional network architectures with a distinct advantage.

Among the top architectures (EfficientNet B5, ResNet 50, custom CNN), we employ two additional criteria to determine the optimal network architecture for USV classification. Firstly, the variance observed across ten cross-validation runs provides insight into the robustness of each architecture. In this regard, both EfficientNet B5 and our custom CNN demonstrate superior stability and minimal variance. Secondly, we evaluate the potential for semi-automated classification. Both architectures achieve approximately \(89\,\mathrm{\%} \) classification accuracy with over \(90\,\mathrm{\%} \) precision. Consequently, EfficientNet B5 and the custom CNN yield comparable results. In conclusion, EfficientNet B5 and custom CNN achieve comparable results. Hence, we favor the smaller network, due to its more transparent architectures and the comparatively modest training times. In summary, our results indicate, that CNNs with residual connections are a good architecture for this type of data and we would recommend our custom CNN for further research. 

Let us comment on some further design decisions.
We did extensive experiments with subsampled spectrograms, which lead to some conclusions, that we did not expect at the beginning. E.g. the FNN architectures allows a rather coarse subsampling without loosing accuracy, indeed best results were achieved with a downsampling to $48 \times8$ spectrograms. 

Results were not so drastic for the custom CNN, but also here a rather coarse downsampling in the time variable is acceptable. In contrast, we need a fine discretization in frequency. As an explanation, we assume that different USVs may differ slightly in their dominant frequencies, but the structure of the calls does not vary significantly over its duration. Hence, we need a fine discretization in frequency, but a downsampling in time does not hamper the quality of classification. As an explanation why downsampling is critical for CNNs, let us have a look 
at the saliency maps. This reveals that the CNN looks at the 'skeleton' of the calls, i.e. at the very fine central line in the spectrogram, this is much harder to detect in blurred and downsampled signal.

Also, the decision to separate detection and classification tasks can be questioned. However, this separation followed standard procedure and allows for a separate, focused analysis of the unique challenges in each task. For instance, the data preprocessing requirements differ significantly. Our refined preprocessing pipeline, particularly tailored for the CNN network, emphasizes the critical nature of this initial step.
 
In addition, we would like to comment on some failed approaches. As already mentioned, we experimented with classification schemes reliant on hand-crafted feature vectors. Specifically, we adopted feature vectors utilized in.\cite{Pessoa2022-sy} However, juvenile USVs are constantly evolving, making it harder to classify them using predefined parameters. Despite efforts to enhance the feature vectors by incorporating additional metrics like energy level ratios, we were unable to achieve accuracies surpassing \(80\,\mathrm{\%} \). 
 
We also observed that preprocessing of the data is crucial. The presented results were obtained after substantial testing of the different preprocessing steps.

Training of the networks, as described, used standard concepts such as Adam, but substantial tests runs were needed before finding the optimal setup for the optimization strategy. This was done by trial and error.

Naturally, the presented study on deep learning concepts for USV classification is not complete. There are other recent concepts, which would be worthwhile to explore in this context. E.g. networks including attention mechanisms or contrastive learning for pre-clustering the USV data or recurrent networks, which would allow to feed in the full USV recording and get directly to a classification without a preliminary detection step are all rather novel and potentially advantages concepts. In this sense, our approach of comparing feed forward networks either based on the classical FNN or on CNN concepts is only a natural first step in the direction of developing optimal algorithms for the task at hand.
%words 539

\section{Conclusion}
\label{sec:Conclusion}

The presented research is part of a larger investigation, which aims at evaluating the phenotype of three mouse lines with suspected autism-like behavior. 
As the manual analysis of the full dataset (593 recordings) would have required more than half a year of rather tiring routine work by a skilled expert, the goal was to develop an automated pipeline capable of detecting and classifying USVs. The presented results demonstrate that indeed such an automated analysis is possible in a completely automated setup with a classification accuracy of \(86.79\,\mathrm{\%} \). 

If this level of accuracy is deemed insufficient, we have outlined a procedure to assess the confidence of the classification. This enables us to establish a subset that can be automatically classified with even greater accuracy, thereby significantly reducing the manual effort required to analyze the original dataset.

We also evaluated whether the AI focuses on the essential aspects of the spectrogram and have provided visual explanations of the classification patterns.

\begin{figure}[ht]
    \includegraphics[width=0.5\textwidth]{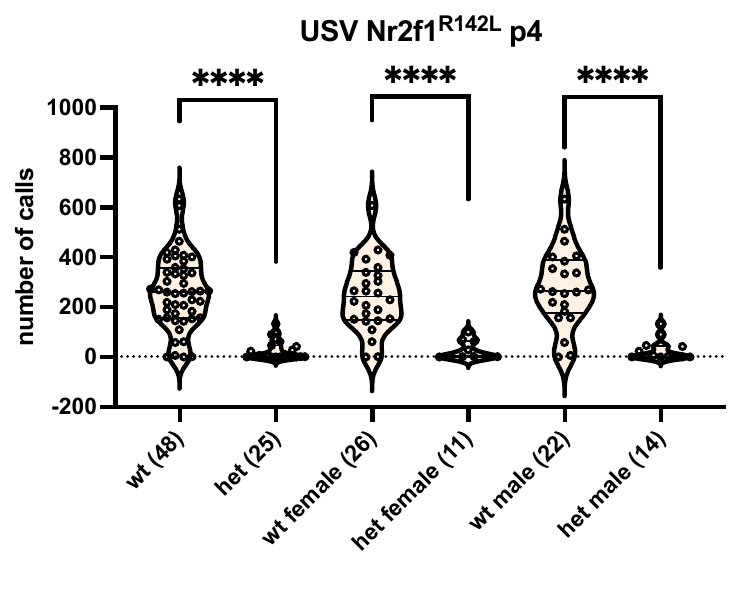}
    \caption{\label{fig:USV_quant_R142L}{Quantitative differences in call number at P4 as detected by our segmentation algorithm.}}
\end{figure}

More importantly, we successfully demonstrated the quantitative and qualitative efficacy of our fully automated pipeline within a research context. In the recordings from our autism-like mouse lines examined in this study, our pipeline revealed significant quantitative differences between wildtype and mutant mice. Specifically, as illustrated in \autoref{fig:USV_quant_R142L}, mutated animals exhibited considerably fewer vocalizations at P4 compared to their wildtype littermates (no difference was detected at P8 and P12). The detection algorithm not only demonstrated high reliability, but also processed a dataset beyond the analytical capacity of a human evaluator.

\begin{figure}[ht]
    \includegraphics[width=0.5\textwidth]{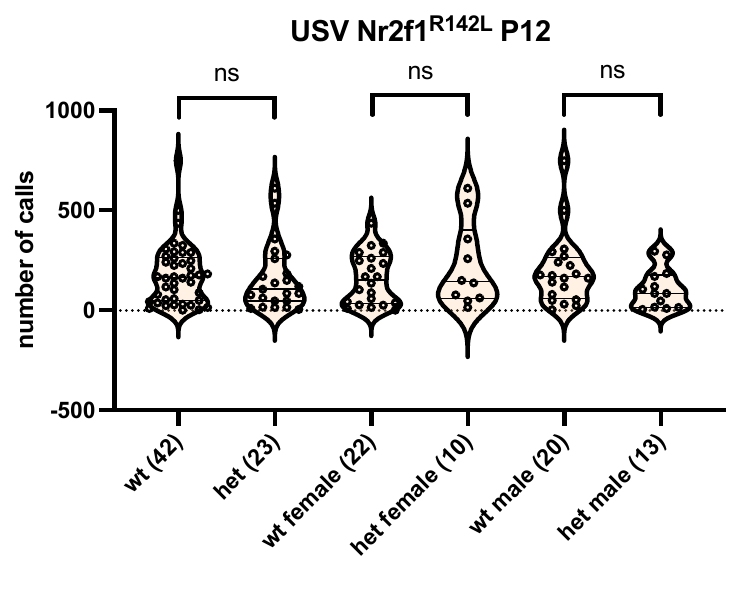}
    \caption{\label{fig:USV_quant_R142Lp12}{No quantitative differences in the call number ar P12, as detected by our segmentation algorithm.}}
\end{figure}

\begin{figure}[ht]
    \includegraphics[width=0.5\textwidth]{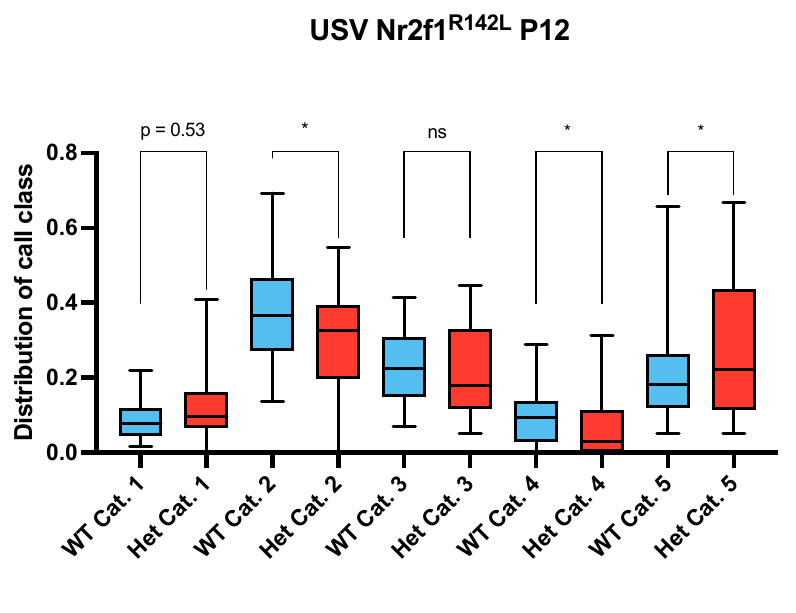}
    \caption{\label{fig:USV_qual_R142Lp12}{Qualitative differences in the call categories at P12, as detected by our classification algorithm.}}
\end{figure}

Moreover, the combined approach of detection and classification enabled us to uncover qualitative differences, such as distinct distributions of calls per class, at developmental stages (P8 and P12) where quantitative analysis alone failed to identify a phenotype, see \autoref{fig:USV_quant_R142Lp12} and \autoref{fig:USV_qual_R142Lp12} for details. Given demonstrated capability of the pipline to analyze subtle differences in pup USVs on a large scale, we will implement the algorithm for further USV analysis at the INBC in Heidelberg.

Over the process of this paper, we have tested different deep learning models in order to analyse which architecture is best for analyzing USVs. For future endeavors we are open to collaborating with other research teams tackling similar tasks in USV classification, thus broadening the scope of the proposed pipeline. We have indicated various avenues where emerging AI concepts like contrastive learning or attention mechanisms could enhance detailed analysis, though the necessity for such advancements remains to be seen. Additionally, we are keen on integrating expert knowledge, such as MEL diagrams, which adjust the frequency scale according to mice perception, potentially enabling further refinement in analysis.

Furthermore, we anticipate that our work will enrich the understanding of both the capabilities and constraints of AI in practical data analysis.
%words 391

\FloatBarrier

\section{Supplementary Material}
\label{sec:Supplementary_material}
%If there is supplementary material for publication, a “SUPPLEMENTARY MATERIAL” section should be added to the manuscript document before the Acknowledgments section.

%% before appendix (optional) and bibliography:
\begin{acknowledgments}
\label{sec:Acknowledgements}
%This research was supported by  ...

We acknowledge the expert advice of Claudia Pitzer and Barbara Kurpiers from the Interdisciplinary Neurobehavioral Core in Heidelberg. R. Herdt is funded by the Deutsche Forschungsgemeinschaft (DFG, German Research Foundation) - project number 459360854 (DFG FOR 5347 Lifespan AI). P. Maass acknowledges the financial support by the Federal Ministry of Education and Research (BMBF) within the T!Raum project "MOIN - MUKIDerm".

\end{acknowledgments}

\section{Author Declarations}
\label{sec:Author_declarations}
The authors have no conflicts of interest to declare.
This study was approved by the Governmental Council Karlsruhe (Project Number G172/21).
The co-authors Herdt, Kinzel, Maaß and Walther contributed equally to this work. The authors Maass and Schaaf are the last authors. 

\section{Data Availability}
\label{sec:Data_Avaliability}
%TODO
The data that support the findings of this study are openly available in [repository name] at http://doi.org/[doi], reference number [reference number].

\section{Appendixes}
\label{sec:Appendixes}
%If you have Figures of tables in the appendix please number them continuous numerical order from all the other Figures and Tables in the manuscript. (Do not namen them Figure A1, Figure A2, Table A1, Table A2, etc.)

%appendix is included in the in the first pdf. Supplementary materials are not

\begin{table}[tb]
  %created 2024-03-02
  %models: ['Y-Net-F_48x8_v1.0_m0', 'Y-Net-F_48x8_v1.0_m1', 'Y-Net-F_48x8_v1.0_m2', 'Y-Net-F_48x8_v1.0_m3', 'Y-Net-F_48x8_v1.0_m4', 'Y-Net-F_48x8_v1.0_m5', 'Y-Net-F_48x8_v1.0_m6', 'Y-Net-F_48x8_v1.0_m7', 'Y-Net-F_48x8_v1.0_m8', 'Y-Net-F_48x8_v1.0_m9']
  %sets: ['Testset_m0', 'Testset_m1', 'Testset_m2', 'Testset_m3', 'Testset_m4', 'Testset_m5', 'Testset_m6', 'Testset_m7', 'Testset_m8', 'Testset_m9']
  \caption{Results of the FNN shown per class, evaluated on the 10 fold cross validation test data.}\label{tab:classes-metrics}
  %\begin{ruledtabular}
  \resizebox{0.5\textwidth}{!}{%
  \begin{tabular}{lccccc}
  \hline
  \hline
    Class & Precision & Recall & Specificity & F1 score & Accuracy\\

    \hline
    all & \( 78.14 \pm 2.48 \) & \( 77.35 \pm 2.55 \) & \( 90.49 \pm 1.14 \) & \( 77.14 \pm 2.61 \) & \( 88.75 \pm 1.23 \)\\
    1 & \( 60.38 \pm 7.98 \) & \( 63.41 \pm 6.65 \) & \( 93.60 \pm 2.25 \) & \( 61.58 \pm 6.69 \) & \( 89.67 \pm 2.45 \)\\
    2 & \( 80.36 \pm 3.62 \) & \( 84.76 \pm 4.26 \) & \( 83.95 \pm 2.65 \) & \( 82.40 \pm 2.66 \) & \( 84.25 \pm 1.98 \)\\
    3 & \( 76.10 \pm 9.14 \) & \( 65.99 \pm 8.33 \) & \( 96.30 \pm 1.62 \) & \( 70.47 \pm 7.75 \) & \( 91.80 \pm 1.97 \)\\
    4 & \( 89.05 \pm 7.72 \) & \( 57.49 \pm 11.60 \) & \( 99.46 \pm 0.34 \) & \( 69.39 \pm 10.17 \) & \( 96.03 \pm 1.28 \)\\
    5 & \( 79.48 \pm 6.84 \) & \( 86.90 \pm 2.50 \) & \( 94.52 \pm 1.80 \) & \( 82.82 \pm 3.39 \) & \( 92.95 \pm 1.18 \)\\

    \hline
    \hline
    
  \end{tabular}
  }
  %\end{ruledtabular}
\end{table}

\begin{figure}[ht]
    %% \reprintcolumnwidth is the same in preprint and reprint for
    %% ease of use for authors:
    \includegraphics[width=0.5\textwidth]{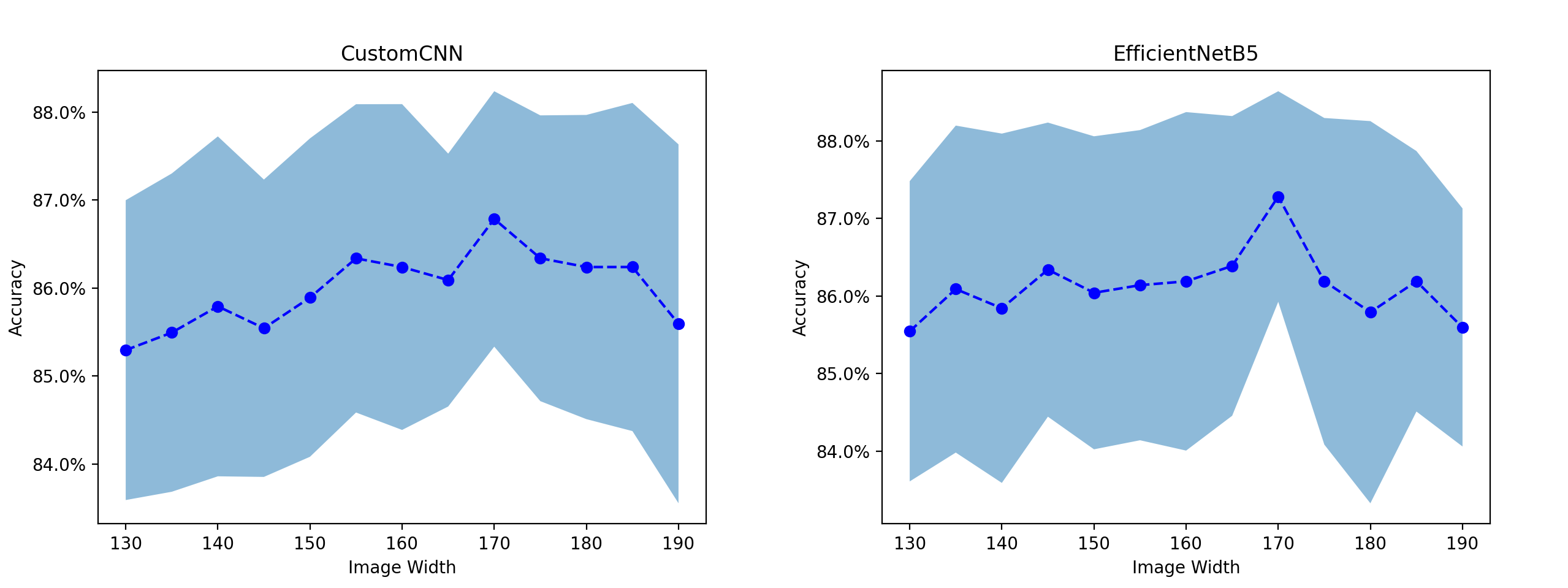}
    \caption{\label{fig:custom_cnn_acc_over_val_size}{Mean accuracy on the validation data of the 10 fold cross validation sets over the duration in ms of the spectrograms. For both the custom CNN and the EfficientNet-B5 the maximium accuracy of 86.79 respectively 87.28 is reached at a size of 170(ms), whereas the training size was 150(ms).}}
    \end{figure}

\begin{table}[tb]
    \caption{\label{tab:baselining_results} Accuracy of the custom CNN when 'removing' inputs, evaluated on the 10 fold cross validation test data.}

    %\begin{ruledtabular}
    \resizebox{0.5\textwidth}{!}{%
    \begin{tabular}{l|cccc}
    \hline
    \hline
     Model & Original & Remove smooth spec & Time feature to avg & Remove db spec \\
    \hline
    Accuracy &  $86.79 \pm 1.45$ & $86.54 \pm 1.52$ & $82.76 \pm 2.19$ & $38.91 \pm 11.88$ \\
    \hline
    \hline
    \end{tabular}
    }
    %\end{ruledtabular}
\end{table}

\subsection{Channel Visualization}
\label{sec:Channel_visualization}

First we compute the channel visualizations of the different layers. These visualizations are computed by activation maximization \cite{ActiviationMaximization} while utilizing transformation robustness as in \cite{olah2017feature}.\\

\begin{algorithm}[ht]
    \label{alg:channel_visualization}
    \caption{Visualize channel c from layer X of the model.}
        \begin{algorithmic}
            \STATE {\bfseries Input:} model from layers 0 to X $F_X$
            \STATE $z^{(0)} \sim \mathcal{N}(0, 0.05)$
            \FOR{$i \gets 0$ to $n$}
                \STATE $\text{loss} \gets \nabla_{z^{(i)}} [F_X(g(z^{(i)}))]_c$ (*)
                \STATE backpropagate loss
                %\STATE optimizer does update step on $z'_i$
                \STATE update $z^{(i)}$
                \ENDFOR
        \end{algorithmic}
    \end{algorithm}
    
    We synthesize those channel visualizations using iterative gradient descent, as shown in Figure \ref{alg:channel_visualization}.
    As optimizer we use Adam with a learning rate of 0.05.\cite{kingma_adam_2015}
    We start from random gaussian noise, and then iteratively update the image $z^{(i)}$ to maximize activation of the channel $c$.
    To reduce noise and make the visualizations more interpretable, we use transformation robustness as in.\cite{olah2017feature}
    This is depicted by the function $g$ in (*) in \ref{alg:channel_visualization}.
    We randomly move the image by up to one pixel and randomly scale it between 0.9 and 1.1.
    One problem with the generation of those visualizations is getting stuck in the initial initialization, if the initial image is not activating the channel $c$ (the channel $c$ being zero everywhere and therefore the gradient being zero and the image will never update).
    To avoid this problem, for the first 16 iterations we change the backward pass of the final ReLU layer in $F_X$ to directly return the incoming gradient (i.e. we allow the gradient backwards even if the activation was zero in the forward pass).
    Also we only optimize the spectrogram, we keep the time feature fixed at the mean time of all the calls.

\begin{figure*}[ht]
    %% \reprintcolumnwidth is the same in preprint and reprint for
    %% ease of use for authors:
    \includegraphics[width=\textwidth]{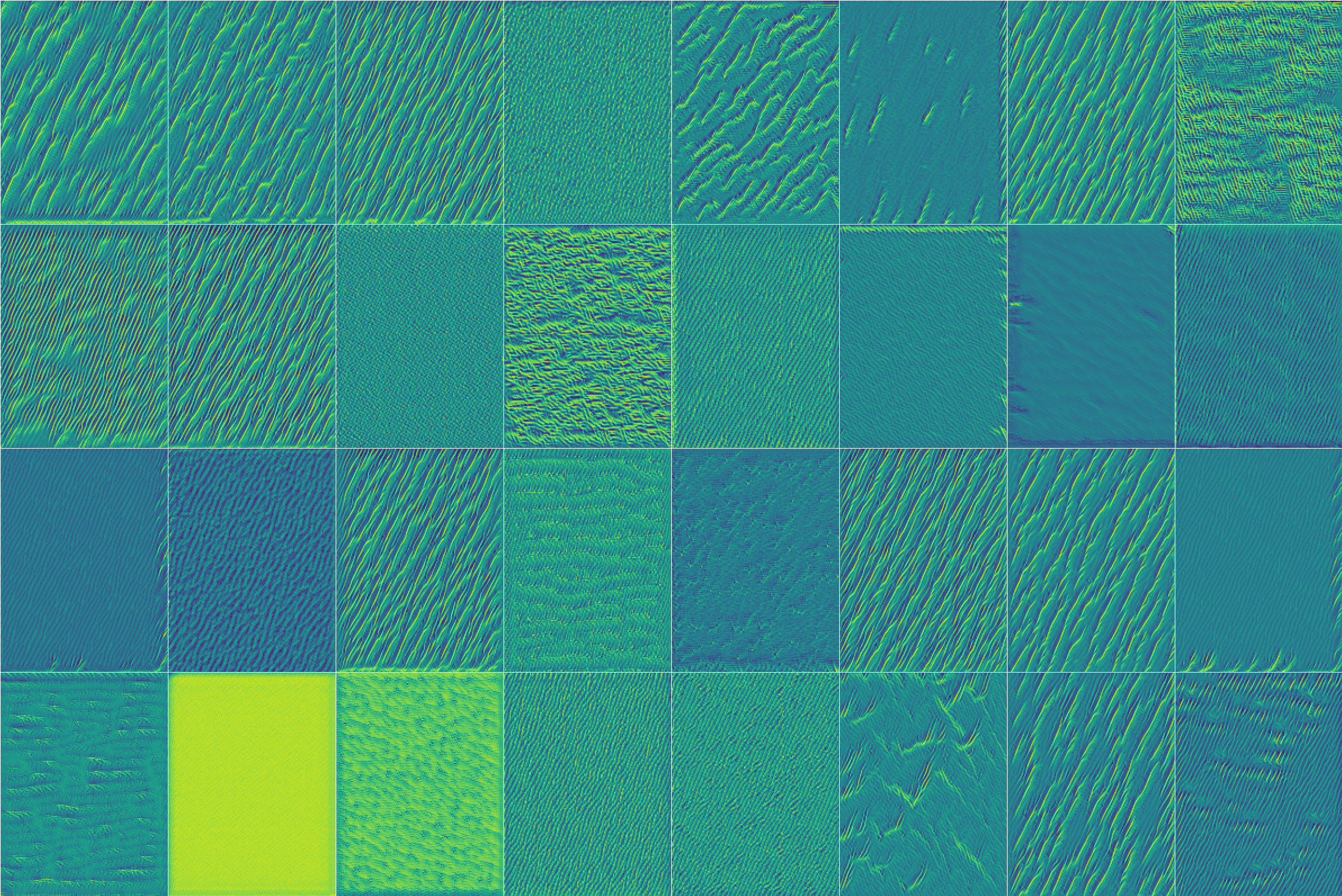}
    \caption{\label{fig:channel_layer0}{}}
    \end{figure*}

\begin{figure*}[ht]
        %% \reprintcolumnwidth is the same in preprint and reprint for
        %% ease of use for authors:
        \includegraphics[width=0.9\textwidth]{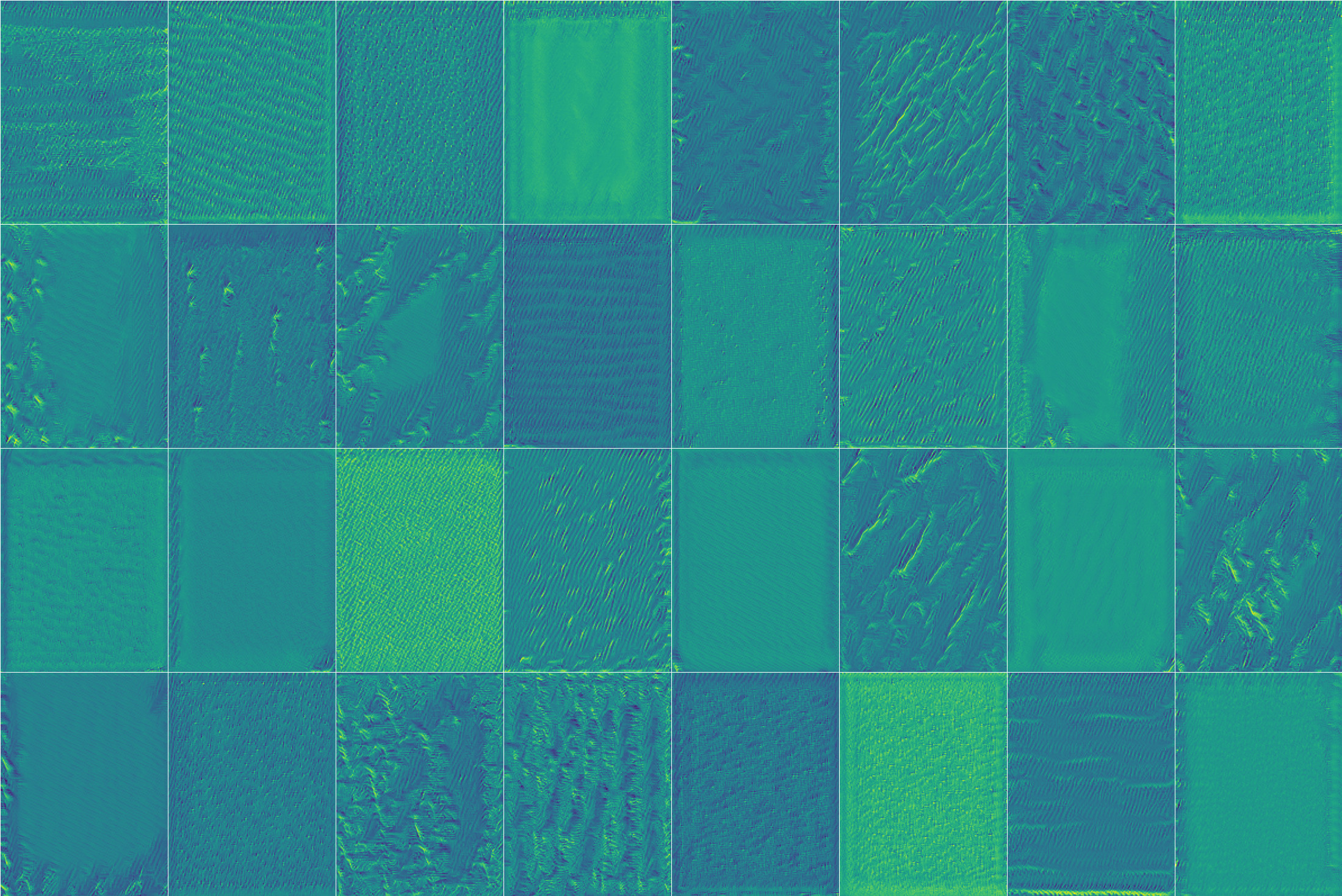}
        \caption{\label{fig:channel_layer1}{}}
        \end{figure*}
\begin{figure*}[ht]
         %% \reprintcolumnwidth is the same in preprint and reprint for
        %% ease of use for authors:
        \includegraphics[width=0.9\textwidth]{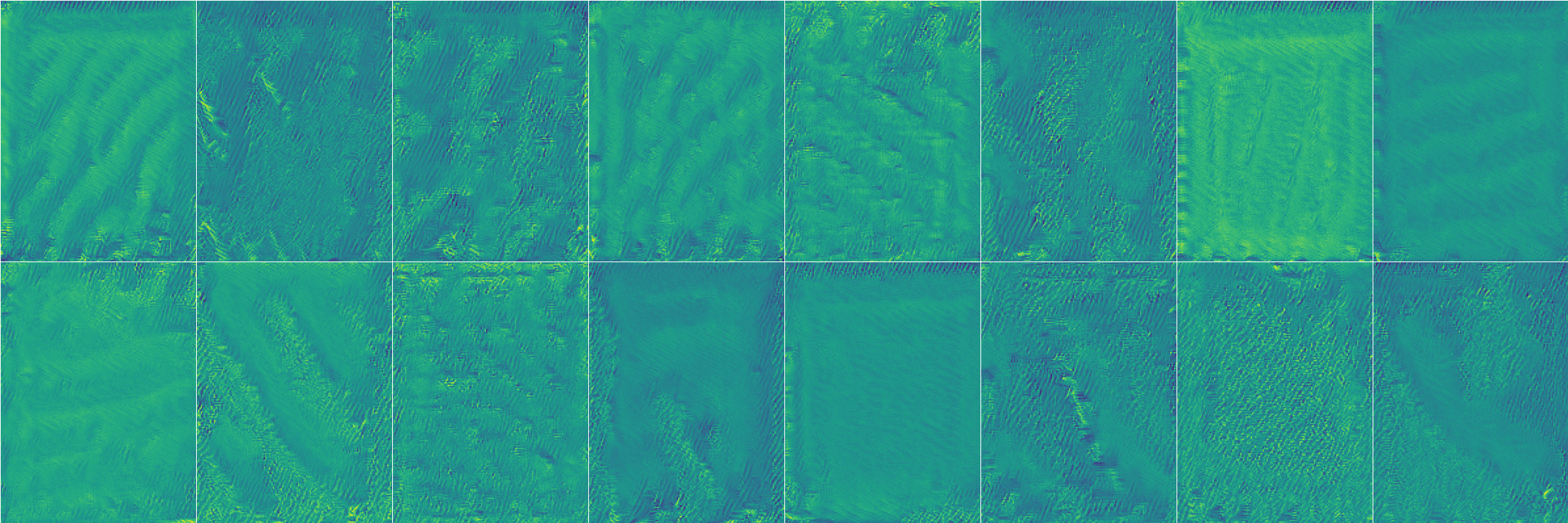}
        \caption{\label{fig:channel_layer2}{}}
        \end{figure*}

\begin{figure}[ht]
        %% \reprintcolumnwidth is the same in preprint and reprint for
        %% ease of use for authors:
        \includegraphics[width=0.5\textwidth]{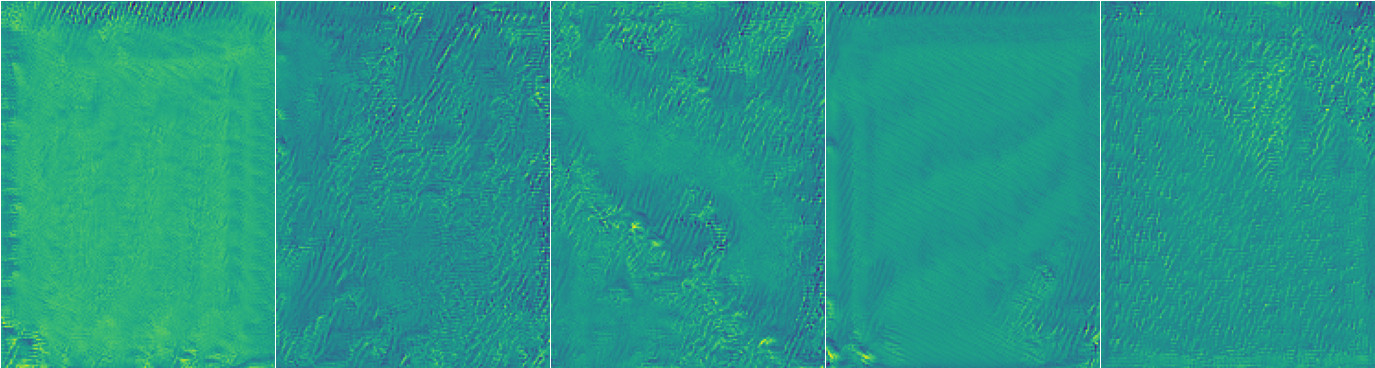}
        \caption{\label{fig:channel_last linear}{}}
        \end{figure}

\subsection{Additional Saliency maps}
\label{sec:Additional_saliency_maps}

We use five samples in SmoothGrad with a noise of standard deviation of 0.1 and 50 samples in Integrated Gradients.
This means, we run Integrated Gradients five times, each time adding random noise with a standard deviation of 0.1 to the image, and return the mean of the five runs as the saliency map.

In Figure \ref{fig:saliency_map} we only plot the dB scale spectrogram, and the saliency map for it, and ignore the smooth spectrogram.
This is due to the insight from Table \ref{tab:baselining_results}, which shows that the model does not rely on the smooth spectrogram for its prediction, but only on the dB scale spectrogram and somewhat on the time feature.

\subsection{Hardware and Software}
\label{sec:Hardware_software}

For the implementations of the neural networks we use the pytorch library, and for setting up multi-GPU training we use the \href{https://github.com/Lightning-AI/lightning}{pytorch-lightning} library.\cite{paszke2019} To compute the saliency maps for interpreting the custom CNN we use the implementations of the captum library.\cite{kokhlikyan2020captum}
We use the implementation of the \href{https://github.com/pytorch/vision}{torchvision library} for the architectures of the ResNet34, ResNet50, EfficientNet-B5 \cite{efficientnetv1} and ViT-B/16 \cite{ViT}.
For the ResNet34, ResNet50 and EfficientNet-B5 we adjust the final fully connected layer (we input the time feature as an additional channel there).
All our experiments regarding the CNNs and the ViT were conducted on a Linux server with 8 Nvidia RTX 2080Ti GPUs. The experiments for the FNN were conducted on Windows with GTX 1080.
%What about fully? Meeting1

%\cite{ActiviationMaximization}
\section{refs}
%bibliographic style or numerical style
\bibliography{usv_paper_bib}
%\bibliography{usv_paper_bib_peter}

\end{document}